\documentclass[12pt]{article}
\usepackage[english]{babel}
\usepackage[latin1]{inputenc}
\usepackage{amsfonts,amssymb,amsmath, epsfig}
\usepackage{color,graphicx,graphics}
\usepackage{amsmath,amstext,amssymb,amsfonts, amscd}
\usepackage[title]{appendix}
\usepackage{hyperref}

\def\Box{\leavevmode\vbox{\hrule
     \hbox{\vrule\kern4pt\vbox{\kern4pt}%
           \vrule}\hrule}}
\def\blackbox{\leavevmode\vrule height 5pt width 4pt depth 0pt\relax}
\def\endproof{\null\hfill {$\blackbox$}\bigskip}

\def\paragraph#1{{\bf #1\ }}

\newtheorem{lemma}{Lemma}[section]

\newtheorem{theorem}[lemma]{Theorem}

\newtheorem{corollary}[lemma]{Corollary}

\newtheorem{definition}[lemma]{Definition}

\newtheorem{proposition}[lemma]{Proposition}

\newtheorem{remark}{Remark}[section]

\newtheorem{assumption}{Assumption}[section]
\title{Evolution of the distribution of wealth in an economic environment driven by local Nash equilibria}
\author{Pierre Degond $^{(1,2)}$, Jian-Guo Liu$^{(3)}$, Christian Ringhofer$^{(4)}$, }
\date{}
\begin{document}

\maketitle

%\vspace{0.5 cm}

\begin{center}
1-Université de Toulouse; UPS, INSA, UT1, UTM ;\\
Institut de Mathématiques de Toulouse ; \\
F-31062 Toulouse, France. \\
2-CNRS; Institut de Mathématiques de Toulouse UMR 5219 ;\\
F-31062 Toulouse, France.\\
email: pierre.degond@math.univ-toulouse.fr
\end{center}

\begin{center}
3- Department of Physics and Department of Mathematics\\
Duke University,
Durham, NC 27708, USA\\
email: jliu@phy.duke.edu
\end{center}

\begin{center}
4- School of Mathematics and Statistical Sciences, Arizona State University, Tempe AZ 85287, USA\\
email: ringhofer@asu.edu
\end{center}

\vspace{0.5 cm}
\begin{abstract}
We present and analyze a model for the evolution of the wealth distribution within a heterogeneous economic environment. The model considers a system of rational agents interacting in a game theoretical framework, through fairly general assumptions on the cost function.  This evolution drives the dynamic of the agents in both wealth and economic configuration variables. We consider a regime of scale separation where the large scale dynamics is given by a hydrodynamic closure with a Nash equilibrium serving as the local thermodynamic equilibrium. The result is a system of gas dynamics-type equations for the density and average wealth of the agents on large scales. We recover the inverse gamma distribution as an equilibrium in the particular case of quadratic cost functions which has been previously considered in the literature. 

\end{abstract}

\medskip
\noindent
{\bf Acknowledgements:} This work has been supported by
KI-Net NSF RNMS grant No. 1107291. JGL and CR are greatful for the opportunity
to stay and work at the Institut de Mathématiques de Toulouse in fall 2012, under the sponsorship of Centre National de la Recherche Scientifique and University Paul--Sabatier. The authors wish to thank A. Blanchet from University Toulouse 1 Capitole for enlighting discussions.


\vskip 0.4cm

%%%%%%%%%%%%%%%%%%%%%%%%%%%%%%%%%%%%%
%%%%%%%%%%%%%%%%%%%%%%%%%%%%%%%%%%%%%
%%%%%%%%%%%%%%%%%%%%%%%%%%%%%%%%%%%%%
%%%%%%%%%%%%%%%%%%%%%%%%%%%%%%%%%%%%%
%%%%%%%%%%%%%%%%%%%%%%%%%%%%%%%%%%%%%
\section{Introduction}
\label{sec:intro}

This paper is concerned with the evolution of wealth in a closed ensemble of agents. The state of each agent is described by two variables. The variable $x$, describes its location in the economic configuration space, as for instance the propensity of an agent to invest \cite{During_Toscani_PhysicaA07}. Therefore, there is a notion of proximity between agents which makes economic interaction (trading) more likely. In addition, the state is described by the wealth $y$ of the agent. The dynamic of these attributes is given by some motion mechanism in the economic configuration variable $x$ and by the exchange of wealth (trading) in the wealth variable $y$.

The basic equation considered in this paper is of the form
\begin{equation}
\partial_t f (x,y,t)+\partial_x ( f \, V(x,y) ) = \partial_y (f \, \partial_y \Phi_f) + d \, \partial^2_y (y^2 f)  ,
\label{intro1}
\end{equation}
where $f(x,y,t)$ is the density of agents in economic configuration space $x$ having wealth $y$ at time $t$.
In the absence of the economic configuration variable $x$, it reduces to a model originally proposed by Bouchaud and M\'ezard \cite{Bouchaud_Mezard_PhypsicaA00} (see also \cite{Cordier_etal_JSP05}), namely
\begin{equation}
\partial_t f (y,t) = \partial_y (f \, \partial_y \Phi_f) + d \, \partial^2_y (y^2 f)  .
\label{intro1_2}
\end{equation}
In this model, $\Phi_f$ is an interaction  `potential', which in the mean-field theory, is functionally dependent on the density $f(x,y,t)$. In \cite{Bouchaud_Mezard_PhypsicaA00}, it describes the resultant of pairwise interactions proportional to the quadratic distance between the wealth of the two agents. The goal of the present paper is to extend this framework to general potentials. The second term at the right hand side of (\ref {intro1_2}) models the uncertainty and has the form of a diffusion operator corresponding to the geometric Brownian motion of economy and finance, with variance $2dy^2$ quadratic in $y$. The justification of this operator can be found in \cite{Oksendal_Springer}. In this model, the total wealth $\int_0^\infty y f(y,t) \, dy$ is preserved. This case is referred to as a ``conservative economy". The Bouchaud and M\'ezard model \cite{Bouchaud_Mezard_PhypsicaA00} can be recovered in the small exchange of wealth limit from the Boltzmann like model of \cite{Cordier_etal_JSP05}. 

In \cite{During_Toscani_PhysicaA07}, D\"uring \& Toscani have introduced an economic configuration variable $x$ to the model of \cite{Bouchaud_Mezard_PhypsicaA00, Cordier_etal_JSP05} and considered Eq. (\ref{intro1}) in the case of a quadratic potential. In their work, the $x$ variable is a propensity to trade. In \cite{During_Toscani_PhysicaA07} as well as in the present model, the position in economic configuration affects the trading behavior of the agent through the dependence of $\Phi_f$ upon $x$. Conversely, the evolution of the agents' wealth $y$ triggers some movement in the economic configuration variable $x$. This evolution is modeled by the quantity $V(x,y)$ which describes how fast the agents move in configuration space (i.e. is the `velocity' in economic configuration). The dependence of $V$ upon both economic configuration $x$ and wealth $y$ expresses that the evolution of the agents in these two variables are coupled. 

In \cite{During_Toscani_PhysicaA07}, the first two moments of the wealth distribution function $f$ are considered. These moments are the density of agents $\rho(x,t)$ and the density of wealth $\rho \Upsilon (x,t)$ defined by:
\begin{equation}
\rho (x,t)=\int f(x,y,t)\ dy, \qquad
\rho \Upsilon (x,t)=\int yf(x,y,t)\ dy \ ,
\label{eq:def_moments}
\end{equation}
(where $\Upsilon$ is the mean wealth) and are functions of the economic configuration $x$ and the time $t$. Taking the moments of (\ref{intro1}) and closing the resulting moment system by using the equilibrium distribution of \cite{Bouchaud_Mezard_PhypsicaA00, Cordier_etal_JSP05}, they find a system of conservation equations for $\rho$ and $\rho Y$ in configuration space $x$ and time $t$, which resembles the gas dynamics equations. 

The first goal of the present paper is to propose an extension of \cite{Bouchaud_Mezard_PhypsicaA00, Cordier_etal_JSP05, During_Toscani_PhysicaA07} by considering arbitrary potentials (and not only quadratic ones). We find that the equilibria cannot be given explicitly, by contrast to the previous literature where they could be expresed in terms of an inverse gamma distribution. Rather, they are found through the resolution of a fixed point equation. If multiple solutions to this fixed point equation exist, corresponding to multiple stable equilibria, this indicates that phase transitions in the wealth distribution are possible. However, we leave the question of the existence and enumeration of the solutions to the fixed point equation to future work.  

The second goal of the paper is to derive the moment equations which provide the dynamics of $\rho$ and $\rho Y$ in configuration space $x$ and time $t$, using the equilibria found from the resolution of the above mentioned fixed point problem. This is done in the same way as in \cite{During_Toscani_PhysicaA07} except for the following differences. First, the conditions of scale separation where this derivation is justified are stated explictly. Second, we give a proof that the density and wealth density are the only quantities which are preserved by the interaction. In kinetic theoretical terms, this means that the only `collision invariants' are linear combinations of $1$ and $y$. This result is rigorously proved here in the case of a quadratic potential considered before \cite{Bouchaud_Mezard_PhypsicaA00, Cordier_etal_JSP05, During_Toscani_PhysicaA07}. The proof relies on a Poincaré inequality which follows from \cite{Benaim_Rossignol_arxiv, Benaim_Rossignol_AnnIHPPS08}. At this point, it is only a conjecture in the case of a general potential. 

The third goal of this paper is to relate the setting of \cite{Bouchaud_Mezard_PhypsicaA00, Cordier_etal_JSP05, During_Toscani_PhysicaA07} to game theory. In the present work, the potential can be viewed as the cost function in a non-atomic non-cooperative anonymous game with a continuum of players \cite{Aumann_Econometrica64, MasColell_JMathEcon84, Schmeidler_JStatPhys73, Shapiro_Shapley_MathOperRes78}, also known as a Mean-Field Game \cite{Cardaliaguet_NotesCollegeFrance12, Lasry_Lions_JapanJMath07}. This cost function arises from the sum of pairwise interactions. For each pair of players, the equilibrium reached under this interaction corresponds to wealth difference being at one of the mimima of this cost function. We note that this model only considers the exchange of money and does not keep track of the goods and services traded. Therefore, this game does not mean that each players wishes to share some of its wealth with the trading partner. Rather, the utility of the exchange is to maximize the economic action resulting in the optimal exchange of goods and services. Within this framework, the dynamic of agents following these strategies can be viewed as given by the following game: each agent follows what is known as the best-reply strategy, that is tries to minimize the cost function with respect to its wealth variable, assuming that the other agents do not change theirs. In order to do so, each player moves in the $y$ direction along the negative gradient of the cost function~$\Phi_f$.

The passage to the macroscopic scale is done by assuming that the spatio-temporal scale of the economic interaction (the dynamic in the $y$-direction) is fast compared to the spatio-temporal scale of the motion in the economic configuration space (i.e. the $x$ variable).  The analysis of such scale separation problems is usually done by re-scaling $t\rightarrow \frac t \varepsilon ,\ x\rightarrow \frac x \varepsilon $, while leaving $y$ unchanged where $\varepsilon \ll 1$ is a small parameter characterizing the ratio of the microscopic units to the macroscopic ones. Typically, here, the microscopic time unit is the inverse of the trading frequency, while the macroscopic one is the characteristic time of the evolution in the economic configuration space. Similar considerations would hold for the spatial units.

This rescaling yields the following perturbation problem:
\begin{equation}
\varepsilon \big( \partial_t f (x,y,t)+\partial_x(f \, V(x,y)) \big)= Q(f),
\label{intro1_1}
\end{equation}
with the operator $Q(f)$ given by
\begin{equation}
Q(f)(y)=\partial_y (f \, \partial_y \bar \Phi_f ) + d \, \partial^2_y (y^2 f)  ,
\label{intro2}
\end{equation}
where $\bar \Phi_f$ is a localized version of the cost function $\Phi_f$. The localized cost function $\bar \Phi_f$ is such that its evaluation at a given point $(x,t)$ only depends on the values of $f$ at the same point. This localization is obtained under the assumption of scale-separation, i.e. that trading interactions are much more likely within agents belonging to the same economic neighborhood $x$ at a given time $t$. For this reason, we will denote it by $\bar \Phi_{\rho(x,t), \nu_{x,t}}(y)$, where 
\begin{eqnarray*}
&&\hspace{-1cm}
\rho(x,t) = \int_{y \in {\mathbb R}_+} f(x,y,t) \, dy, \qquad \nu_{x,t}(y) = \frac{1}{\rho(x,t)} f(x,y,t).
\end{eqnarray*}
respectively denote the local density of agents at $(x,t)$ and the conditional probability of presence of the agents, conditioned on fixing $(x,t)$. Writing $\Bar \Phi = \bar \Phi_{\rho(x,t), \nu_{x,t}}(y)$ expresses that $\bar \Phi$ evaluated at $(x,t)$ depends on the scalar quantity $\rho(x,t)$ and is functionally dependent on the probability $\nu_{x,t}$.

In the limit $\varepsilon \to 0$, one is led to look for the solutions of $Q(f) = 0$. These solutions, known as the local equilibria, provide the formal limit of the solution $f^\varepsilon$ to (\ref{intro1_1}) when $\varepsilon \to 0$. Along with the interpretation given in \cite{Degond_etal_preprint13_2}, we show that these local equilibria can be interpreted as Nash equilibria of the mean-field game defined above.  So, Eq. (\ref {intro1}) can be interpreted as the mean field limit of a system of agents whose dynamic in the wealth variable $y$ is driven by the best-reply strategy, i.e. is a march towards the Nash equilibrium along the gradient of the cost function. This motion is supplemented with the standard geometric Brownian motion of economy and finance.

The Nash equilibria can be expressed in the form of Gibbs distributions $M_\Xi(y)$:
\begin{eqnarray}
M_\Xi(y) &=& \frac{1}{Z_\Xi} \exp \big( - \frac{\Xi(y)}{d} \big), \qquad Z_\Xi = \int_{y \in {\mathbb R}_+} \exp \big( - \frac{\Xi(y)}{d} \big) \, dy.
\label{eq:intro_Nash}
\end{eqnarray}
where the function $\Xi(y)$ is associated to the cost function $\bar \Phi_{\rho, \nu}$. For $M_\Xi$ to be a local equilibrium, $\Xi$ must satisfy a fixed point equation expressing the Nash equilibrium property. Since $\bar \Phi_{\rho(x,t), \nu_{x,t}}$ depend locally on $(x,t)$, so do the function $\Xi = \Xi_{x,t}(y)$. The importance of the Gibbs distribution in the statistical physics of money and wealth is reviewed in~\cite{Yakovenko_Rosser_RevModPhys09}. 

The large scale limit i.e. the limit $\varepsilon \rightarrow0$ after rescaling $t\rightarrow\frac t \varepsilon ,\ x\rightarrow\frac x \varepsilon $ yields the gas-dynamic type system of evolution equations for the macroscopic quantities (the moments), i.e. for the local density $\rho(x,t)$ and average wealth $\Upsilon(x,t)$ defined at (\ref{eq:def_moments}). 
The macroscopic equations are of the form
\begin{eqnarray}
&&\hspace{-1cm}
\partial_t \rho (x,t) + \partial_x (\rho \, u(x; \Xi_{x,t} ))=0,
\label{intro3} 
\\
&&\hspace{-1cm}
\partial_t (\rho \Upsilon)  (x,t) +  \, \partial_x (\rho {\mathcal E}(x; \Xi_{x,t} ) ) =0
\ .
\label{intro3_1}
\end{eqnarray}
Here the macroscopic velocity $u(x;\Xi_{x,t})$ and wealth velocity ${\mathcal E}(x;\Xi_{x,t})$ are functions of $x$ and are functionally dependent on $\Xi_{x,t}$. They are computed via closure of the moments of equation (\ref {intro1}) using the Nash equilibrium (\ref{eq:intro_Nash}) as a closure. For the case of a general trading interaction potential, it is not known wheter system (\ref{intro3}), (\ref{intro3_1}) forms a closed system of equations, i.e. in other words, if the knowledge of $\rho$ and $\Upsilon$ suffices to reconstruct $\Xi$ in a unique way. This point is left to future work. 

However, in the case of a quadratic trading interaction \cite{Bouchaud_Mezard_PhypsicaA00, Cordier_etal_JSP05, During_Toscani_PhysicaA07}, it can be shown that it is actually the case and we can write $u(x;\Xi_{x,t}) = u(x;\Upsilon(x,t))$,  ${\mathcal E}(x; \Xi_{x,t})= {\mathcal E}(x;\Upsilon(x,t))$, which makes system (\ref{intro3}), (\ref{intro3_1}) a closed system of equations. Under suitable assumptions on $u$ and ${\mathcal E}$, it can be shown that this system is hyperbolic. Therefore, it is (at least locally in time) well-posed and shares similar features as the classical system of compressible gas dynamics equations. This system has already been proposed in \cite{During_Toscani_PhysicaA07}. However, we prove that the conservation of density and wealth density are the only conserved quantities. The proof of this fact relies on a Poincaré inequality which can be derived from \cite{Benaim_Rossignol_arxiv, Benaim_Rossignol_AnnIHPPS08}. This is an important step towards the rigorous proof of the convergence of the solutions of (\ref{intro1_1}) to those of (\ref{intro3}), (\ref{intro3_1}) in the $\varepsilon \to 0$ limit. 

We now give a few examples of potential applications of the present work, beyond the case considered in \cite{During_Toscani_PhysicaA07}, where $x$ has the meaning of a propensity to trade. In the first example, the economic configuration variable $x$ coincides with the geographical position~$x$. Here for instance, we are considering different countries, or different areas of the same country, which have different wealth distributions. It is a documented fact \cite{Garip_report, McKenzie_Rapoport_JDevEcon07} that spatial inhomogenities of the wealth distribution trigger migrations, people from areas where the lower range of the wealth distributions is thick tending to migrate towards areas where the higher range of the wealth distribution is more populated. Obviously, in return, these migrations affect the shape of the wealth distribution in the concerned areas. In this context, the model can be used to provide a spatially continuous description of wealth distribution inhomogeneities and of their evolution in the large time due to migrations.

The second example has to do with the connection between wealth and social status which has been studied in e.g. \cite{Corneo_Jeanne_ScandJEcon01, Robson_Econometrica92, Weiss_etal_EurEconRev98}. Social status has an obvious relation to wealth. However, this relation may not be straightforward, as social status may also be influenced by other factors such as professional or marital status and power. Obviously, a higher social status increases the opportunities to make money and conversely, money may buy some factors which influence social status. Therefore, it is interesting to study how differences between wealth distributions in different social strata may trigger movement of agents between these strata in the long term and how, in return, these movements influence the wealth distribution in each strata.

Lastly, the third example pertains to the connection between wealth and education or cultural level. Studies about this connection can be found e.g. in \cite{Fershtman_etal_EconJ93, Galor_Zeira_RevEconStud93}.  This is very similar to the previous example as a higher education or cultural level increases the opportunities for a high trading activity. Conversely, an increased wealth due to a higher economic activity provides the resources and incentives for training and participation in cultural activities. Therefore, like in the previous examples, our methodology could be used to study the long term evolution of the distribution of education or cultural level due to inhomogeneities in the wealth distribution among the various cultural or educational strata.

This paper is organized as follows. In Section \ref{sec:finance}, we present the multi-agent model for the dynamics of $N$  agents. We assume the existence of a mean field limit for $N\rightarrow \infty $, yielding formally the Fokker-Planck equation (\ref {intro1}) for the effective single agent density $f(x,y,t)$. Section \ref{sec:homogeneous} is devoted to the spatially homogeneous case, where $f$ is independent of the economic configuration variable $x$. Its main purpose is to provide the fixed point equation whose resolution yields the equilibrium distribution of $Q$ in (\ref {intro2}). We show that the equilibrium distribution is actually a Nash equilibrium, i.e. no player can improve on the cost function by choosing a different direction in $y$. In Section \ref{sec:hydro} we consider the inhomogeneous case and close the moments of the kinetic equation (\ref {intro1}) using the equilibrium density from Section 3. Finally, we conclude by drawing some perspectives in section \ref{sec:conclu}.

%%%%%%%%%%%%%%%%%%%%%%%%%%%%%%%%%%%%%
%%%%%%%%%%%%%%%%%%%%%%%%%%%%%%%%%%%%%
%%%%%%%%%%%%%%%%%%%%%%%%%%%%%%%%%%%%%
%%%%%%%%%%%%%%%%%%%%%%%%%%%%%%%%%%%%%
%%%%%%%%%%%%%%%%%%%%%%%%%%%%%%%%%%%%%
\setcounter{equation}{0}
\section{The model and its mean-field limit}
\label{sec:finance}

\subsection{A dynamic agent model for wealth distribution}
\label{subsec:agent}

We consider a set of $N$ market agents. Each agent, labeled $j$, is endowed with two variables: its wealth $Y_j \in {\mathbb R}_+$ and a variable $X_j \in {\mathbb R}$ which characterizes its economic configuration, i.e. the category of agents it ususally interacts with. We ignore the possibility of debts so that we take $Y_j \geq 0$.

The dynamics of the agents is an elaboration of a model proposed in \cite{Bouchaud_Mezard_PhypsicaA00}. It is written:
\begin{eqnarray}
&&\hspace{-1cm}
\dot X_j = V(X_j(t), Y_j(t)),
\label{eq:dd_wealth_X} \\
&&\hspace{-1cm}
dY_j = - \frac{1}{N} \sum_{k\not = j} \xi_{jk} \, \Psi(|X_j - X_k|) \, \, \partial_Y \phi (Y_j - Y_k) \, dt + \sqrt{2d} \, \, Y_j  \, dB_t^j.
\label{eq:dd_wealth_Y}
\end{eqnarray}
The first term at the right-hand side of (\ref{eq:dd_wealth_Y}) describes economic interactions such as trading. The quantity $\phi$ is a trading interaction potential which describes how the trading activity depends on the difference of wealth between the trading agents. In \cite{Bouchaud_Mezard_PhypsicaA00}, the trading activity is proportional to the difference of wealth, which means that $\phi$ is a quadratic function. This simple assumption is convenient as it gives rise to explicit formulae for the equilibria (see below). However, in the perspective of making this theory more quantitative, it is desirable to investigate more general forms of trading interactions. Here, at this point, we make no assumption on the form of the trading interaction $\phi$. The weight $\xi_{jk} \, \Psi(|X_j - X_k|)$ is the trading frequency between agents $j$ and $k$ and depends on the distance in economic configuration space between the two agents. We assume that $\xi_{jk}$ is symmetric: $\xi_{jk} = \xi_{kj}$ and that $\Psi$ is normalized:
\begin{eqnarray}
&&\hspace{-1cm}
\int_{x \in {\mathbb R}} \Psi (|x|) \, dx = 1.
\label{eq:dd_normal}
\end{eqnarray}
The second term at the right-hand side of (\ref{eq:dd_wealth_Y}) is the classical geometric Brownian motion of finance and $\sqrt{2d}$ is the volatility. The quantities $B_t^j$ denote independent Brownian motions. Eq. (\ref{eq:dd_wealth_X}) describes how fast the agent evolves in the economic configuration space as a function of its current wealth and current economic configuration and $V(x,y)$ is a measure of the speed of this motion.

In this paper, we make the following important assumption:

\begin{assumption}
The function $Y \in {\mathbb R} \to \phi(Y) \in {\mathbb R}$ is $C^2$ and even.
\label{ass:even}
\end{assumption}

Under this assumption, the first term of (\ref{eq:dd_wealth_Y}) conserves the total wealth. Indeed, the rates of evolution of the wealth of agent $j$ in its interaction with agent $k$ and conversely, of agent $k$ in its interaction with agent $j$ are opposite and add up to zero. The stochastic differential equation (\ref{eq:dd_wealth_Y}) should be understood in the It\`o sense. This is a difference with \cite{Bouchaud_Mezard_PhypsicaA00} where the Stratonovich sense was used. However, upon a simple conversion involving the addition of a term $Y_j \, d \, dt$, the models are the same in the case of a quadratic trading interaction potential $\phi$. In this case, Eq. (\ref{eq:dd_wealth_Y}) is homogeneous of degree one with respect to the variable $Y_j$, which means that the unit of wealth in this case can be arbitrary. However, this homogeneity is lost for a more general trading interaction potential, and later on, we will have to choose a unit of wealth (such as a given monetary unit). 

In this work, we assume that the trading frequency $\xi_{jk}$ is a function of the number of trading agents in the economic neighbourhoods of $j$ and $k$. For instance, we can take
\begin{eqnarray*}
&&\hspace{-1cm}
\xi_{jk} = \xi \big( \frac{\rho^\Psi_j + \rho^\Psi_k}{2} \big)
%\label{eq:alpha_jk}
\end{eqnarray*}
where
\begin{eqnarray*}
&&\hspace{-1cm}
\rho^\Psi_j = \frac{1}{N} \sum_{\ell \not = j} \, \Psi(|X_\ell - X_j|) .
%\label{eq:rho_Psi_j}
\end{eqnarray*}
For the simplicity of notations, we have assumed that the 'counting function' $\Psi$ is the same as in (\ref{eq:dd_wealth_Y}) but this restriction in unessential.

We introduce the notations $\vec{X}(t) =(X_1, \ldots, X_N)$, $\vec{Y}(t) =(Y_1, \ldots, Y_N)$ and $\hat Y_j = (Y_1, \ldots, Y_{j-1},$ $ Y_{j+1}, \ldots, Y_N)$ (note that in game theory, $\hat Y_j$ is often denoted $Y_{-j}$). We also write ${\vec Y} = (Y_j, \hat Y_j)$ by abuse of notation. Eq. (\ref{eq:dd_wealth_Y}) can be recast in the following form:
\begin{eqnarray}
&&\hspace{-1cm}
dY_j = - \partial_{Y_j} \Phi^N(\vec{X}, Y_j, \hat Y_j, t) \, dt +  \sqrt{2d} \,\,Y_j \,   dB_t^j.
\label{eq:dd_wealth_Y_2}
\end{eqnarray}
where the cost function $\Phi^N(\vec{X}, \vec{Y}, t)$ is given by
\begin{eqnarray}
&&\hspace{-1cm}
\Phi^N(\vec{X}, \vec{Y}, t) = \frac{1}{N} \sum_{k\not = j} \xi_{jk}(\vec{X}) \, \Psi(|X_j - X_k|) \, \phi(Y_k - Y_j) ,
\label{eq:dd_potential}
\end{eqnarray}
with
\begin{eqnarray*}
&&\hspace{-1cm}
\xi_{jk} (\vec{X}) = \xi \big( \frac{\rho^\Psi_j(\vec{X}) + \rho^\Psi_k(\vec{X})}{2} \big) , \qquad \rho^\Psi_j (\vec{X}) = \frac{1}{N} \sum_{\ell \not = j} \, \Psi(|X_\ell - X_j|) .
\end{eqnarray*}
Therefore, the agents choose the steepest descent direction of their cost function $Y_j \to \Phi^N(\vec{X}, Y_j, \hat Y_j)$ as their action in wealth space. This action is supplemented with a geometric Brownian noise which models volatility. Ignoring these additional effects due to the environment, the agents would eventually, at large times, reach a point of minimum of their cost function. This minimum would then be written
\begin{eqnarray}
&&\hspace{-1cm}
Y_j^N (\vec{X}, \hat Y_j,t) = \mbox{arg} \min_{Y_j \in {\mathbb R}_+} \Phi^N(\vec{X}, Y_j, \hat Y_j, t), \quad \forall j \in \{ 1, \ldots, N\}.
\label{eq:discrete_Nash}
\end{eqnarray}
and corresponds to a Nash equilibrium of the agents. Therefore, the dynamics correspond to a
non-cooperative non-atomic anonymous game \cite{Aumann_Econometrica64, MasColell_JMathEcon84, Schmeidler_JStatPhys73, Shapiro_Shapley_MathOperRes78}, also known as a Mean-Field Game \cite{Cardaliaguet_NotesCollegeFrance12, Lasry_Lions_JapanJMath07}, where the equilibrium assumption is replaced by a time dynamics
describing the march towards a Nash equilibrium. However, the cost function depends on the position of the
agent in economic configuration space and reciprocally, the motion of the agents in economic configuration space depends on their wealth. Consequently, this march towards a local Nash equilibrium may be perturbed by the motion of the agents in economic configuration space. The goal of this paper is to study the large scale dynamics of this system by establishing macrocopic equations in economic configuration space, particularly in the case where the potential is not quadratic (the quadratic case being treated in \cite{During_Toscani_PhysicaA07}). To this aim, we consider a continuous version of this discrete system.

\subsection{Mean-Field limit}
\label{subsec_mean_field}

We introduce the $N$-particle empirical distribution function
$$ f^N(x,y,t) = \frac{1}{N} \sum_{j=1}^N \delta_{X_j(t)}(x) \otimes \delta_{Y_j(t)}(y), $$
and regard $f^N$ as a map from $t \in {\mathbb R}_+$ to $f^N(t) \in {\mathcal P}({\mathbb R} \times {\mathbb R}_+)$, where ${\mathcal P}({\mathbb R} \times {\mathbb R}_+)$ denotes the set of probability measures on ${\mathbb R} \times {\mathbb R}_+$.
We now make the following assumptions: 

\begin{assumption}
In the mean-field limit $N \to \infty$ of the number of agents going to infinity, there exists a one-particle distribution function $f = f(x,y,t)$, which maps $t \in {\mathbb R}_+$ to $f(t) \in {\mathcal P}_{\mbox{\scriptsize ac}}({\mathbb R} \times {\mathbb R}_+)$ where ${\mathcal P}_{\mbox{\scriptsize ac}}({\mathbb R} \times {\mathbb R}_+)$ is the space of probability measures on $({\mathbb R} \times {\mathbb R}_+)$ which are absolutely continuous with respect to the Lebesgue measure on $({\mathbb R} \times {\mathbb R}_+)$, such that
\begin{equation}
f^N \rightharpoonup f, \qquad \mbox{ when } N \to \infty,
\label{eq:weak_cvg}
\end{equation}
in the weak star topology of bounded measures. 
\label{ass:mean-field_1}
\end{assumption}

We also assume that a mean-field cost function exists:

\begin{assumption}
We assume that there exists a map ${\mathcal P}_{\mbox{\scriptsize ac}}({\mathbb R} \times {\mathbb R}_+) \to C^2({\mathbb R} \times {\mathbb R}_+)$, $f \mapsto \Phi_{f}$,   such that, for all trajectories $(X_j(t), Y_j(t))$ satisfying (\ref{eq:weak_cvg}) and $(X(t), Y(t))$ such that $(X_j(t), Y_j(t)) \to (X(t), Y(t))$ when $N \to \infty$, we have
\begin{eqnarray}
&&\hspace{-1.2cm}
\Phi^N \big( X_j(t), \hat X_j(t), Y_j(t), \hat Y_j(t), t \big) \to \Phi_{f(t)} \big( X(t), Y(t)  \big) , \, \, \forall j \in \{ 1, \ldots, N\}, \, \, \forall t \geq 0.
\label{eq:mfl_pot}
\end{eqnarray}
\label{ass:mean-field_2}
\end{assumption}

Under Assumptions \ref{ass:mean-field_1} and \ref{ass:mean-field_2}, and owing to (\ref{eq:dd_potential}), $\Phi_{f(t)}$ is given by:
\begin{eqnarray}
&&\hspace{-1.5cm}
\Phi_{f(t)}(x,y) = \int_{(x',y') \in {\mathbb R} \times {\mathbb R}_+}  \xi \big(\frac{\rho^\Psi(x,t)+ \rho^\Psi(x',t)}{2} \big) \, \Psi(|x'-x|) \nonumber \\
&&\hspace{6.5cm}
 \phi (y' -y) \, f(x', y', t) \, dx' \, dy' ,
\label{eq:mfl_potential}
\end{eqnarray}
with
\begin{eqnarray*}
&&\hspace{-1cm}
\rho^\Psi(x,t) =\int_{(x',y') \in {\mathbb R} \times {\mathbb R}_+} \Psi(|x-x'|) \, \, f(x',y',t) \, dx' \, dy'.
\end{eqnarray*}
Then, in the limit $N \to \infty$, the one-particle distribution function $f$ is a solution of the following Fokker-Planck equation \cite{Bouchaud_Mezard_PhypsicaA00}:
\begin{eqnarray}
&&\hspace{-1cm}
\partial_t f + \partial_x (V(x,y) f) + \partial_y (F_f \, f) = d \, \partial_y^2 \big( y^2 f \big),
\label{eq:mfl_eq_finance}
\end{eqnarray}
where $F_f = F_f(x,y,t)$ is given by
\begin{eqnarray}
&&\hspace{-1cm}
F_f(x,y,t) = - \partial_y \Phi_{f(t)} (x,y) .
\label{eq:mfl_force}
\end{eqnarray}
For short, we will write $\Phi_{f(t)} = \Phi_f$. Thanks to (\ref{eq:mfl_potential}), $F_f$ can be written:
\begin{eqnarray}
&&\hspace{-1.5cm}
F_f(x,y,t) =- \int_{(x',y') \in {\mathbb R} \times {\mathbb R}_+} \xi \big( \frac{\rho^\Psi(x,t)+\rho^\Psi(x',t)}{2} \big) \, \Psi(|x-x'|) \nonumber \\
&&\hspace{6.5cm}
\, \partial_y \phi (y-y') \, f(x',y',t) \, dx' \, dy'.
\label{eq:mfl_force_finance}
\end{eqnarray}
We supplement this equation with the boundary condition
\begin{eqnarray}
&&\hspace{-1cm}
f(x,0,t) = 0, \quad \forall x \in {\mathbb R}, \quad \forall t \in {\mathbb R}_+.
\label{eq:mfl_bc_finance}
\end{eqnarray}
We also provide an initial condition $f(x,y,0) = f_0(x,y)$.

\subsection{Macroscopic scaling}
\label{subsec_macro_scaling}

In order to manage the various scales in a proper way, we first change the variables to dimensionless ones. We introduce $t_0$ and $x_0 = a t_0$ the time and economic configuration space units, with $a$ the typical magnitude of $V$. We scale the wealth variable $y$, by a monetary unit $y_0$. We choose $t_0$ in such a way that the magnitude of  $\Psi$ and $d$ are ${\mathcal O}(1)$. Introducing $\tilde x = x/x_0$, $\tilde t = t/t_0$, $\tilde y = y/y_0$, $\tilde f (\tilde x, \tilde y, \tilde t)= x_0 \, y_0 \, f(x,y,t)$, $\tilde V(\tilde x, \tilde y) = V(x,y) / a$, $\tilde \Psi(|\tilde x- \tilde x'|) = x_0 \Psi(| x- x'|)$, $\tilde \rho^{\tilde \Psi} (\tilde x, \tilde t) = x_0 \rho^\Psi(x,t)$, $\tilde \xi (\tilde \rho) = \xi (\rho) / (a y_0^2)$, $\tilde F_{\tilde f} (\tilde x, \tilde y, \tilde t) = (t_0/y_0) F_f(x,y,t)$, $\tilde \Phi_{\tilde f} (\tilde x, \tilde y, \tilde t) = (t_0/y_0^2) \Phi_f(x,y,t)$, $\tilde \phi (\tilde y) = \phi (y)$, $\tilde d = d t_0$, Eq. (\ref{eq:mfl_eq_finance}) is written:
\begin{eqnarray}
&&\hspace{-1cm}
\partial_{\tilde t} {\tilde f} + \partial_{\tilde x} (\tilde V(\tilde x,\tilde y) {\tilde f}) + \partial_{\tilde y} ({\tilde F}_{\tilde f} \, {\tilde f}) = \tilde d \, \partial^2_{\tilde y} \big(\tilde y^2 \tilde f \big),
\label{eq:mfl_eq_tilde_finance}
\end{eqnarray}
with ${\tilde F}_{\tilde f}$  given by
\begin{eqnarray}
&&\hspace{-1cm}
{\tilde F}_{\tilde f}(\tilde x, \tilde y,\tilde t) = - \partial_{\tilde y} \tilde \Phi_{\tilde f} (\tilde x, \tilde y, \tilde t),
\label{eq:mfl_force_tilde_finance}
\end{eqnarray}
and
\begin{eqnarray}
&&\hspace{-1.5cm}
\tilde \Phi_{\tilde f} (\tilde x, \tilde y, \tilde t) = \nonumber \\
&&\hspace{-1.cm}
= \int_{(\tilde x', \tilde y') \in {\mathbb R} \times {\mathbb R}_+} \tilde \xi \big( \frac{\tilde \rho^{\tilde \Psi}(\tilde x, \tilde t)+ \tilde \rho^{\tilde \Psi}(\tilde x',\tilde t)}{2} \big) \,  \tilde \Psi(|\tilde x- \tilde x'|) 
\, \phi(\tilde y - \tilde y') \, \tilde f( \tilde x', \tilde y', \tilde t) \, d\tilde x' \, d \tilde y',
\label{eq:mfl_pot_tilde_finance} \\
&&\hspace{-1.5cm}
\tilde \rho^{\tilde \Psi}(\tilde x, \tilde t) =\int_{(\tilde x', \tilde y') \in {\mathbb R} \times {\mathbb R}_+} \tilde \Psi(|\tilde x- \tilde x'|) \, \, \tilde f(\tilde x', \tilde y', \tilde t) \, d \tilde x' \, d \tilde y'.
\label{eq:mfl_dens_tilde_finance}
\end{eqnarray}
We note that $\tilde \Psi$ still satisfies the normalization condition (\ref{eq:dd_normal}).

Following the procedure developed in \cite{Degond_etal_preprint13_2}, we introduce the macroscopic scale. We assume that the changes in economic configuration $x$ are slow compared to the exchanges of wealth between agents. Therefore, we change the units of economic configuration space and time to new ones $x'_0 = x_0 / \varepsilon$, $t'_0 = t_0 / \varepsilon$ with $\varepsilon \ll 1$. The parameter $\varepsilon$ is the ratio of the typical time between two economic interactions (which is very short) compared to the time-scale of the evolution of the economic configuration variable (which is large). Therefore, we change the space and time variables to macroscopic variables $\hat x = \varepsilon \tilde x$, $\hat t = \varepsilon \tilde t$ and define $\hat f (\hat x, \tilde y, \hat t) = \varepsilon^{-1} \tilde f (\tilde x, \tilde y, \tilde t)$. We note that, as usual in kinetic theory, we do not scale the variable $\tilde y$. In this change of scale, the integral (\ref{eq:mfl_dens_tilde_finance}) becomes $\int \tilde \Psi(\frac{|\hat x- \hat x'|}{\varepsilon}) \, \, \hat f(\hat x',\tilde y', \hat t) \, d \hat x' \, d \tilde y' = {\mathcal O}(\varepsilon)$. Therefore, we rescale the density $\tilde \rho^{\tilde \Psi}$ to take into account this behavior and let $\hat \rho^{\tilde \Psi} (\hat x, \hat t) = \varepsilon^{-1} \tilde \rho^{\tilde \Psi} (\tilde x, \tilde t)$. We scale the velocity, cost function and force to unity i.e. we let $\hat V(\hat x, \tilde y ) = \tilde V (\tilde x, \tilde y)$, $\hat F_{\hat f} (\hat x, \tilde y, \hat t) = \tilde F_{\tilde f} (\tilde x, \tilde y, \tilde t)$, $\hat \Phi_{\hat f} (\hat x, \tilde y, \hat t) = \tilde \Phi_{\tilde f} (\tilde x, \tilde y, \tilde t)$.

Similar to (\ref{eq:mfl_dens_tilde_finance}), in this rescaling, the integral in (\ref{eq:mfl_pot_tilde_finance}) is ${\mathcal O}(\varepsilon)$. To compensate for it, we assume that $\tilde \xi$ is large, which maintains the trading frequency of order unity. Therefore, we let $\hat \xi (\hat \rho) = \varepsilon \tilde \xi (\tilde \rho)$. The resulting expression of the force term is
\begin{eqnarray}
&&\hspace{-1cm}
{\hat F}_{\hat f}({\hat x},\tilde y,{\hat t}) = - \partial_{\tilde y} \hat \Phi_{\hat f} (\hat x, \tilde y, \hat t),
\label{eq:mfl_force_tilde_finance_2}
\end{eqnarray}
with
\begin{eqnarray}
&&\hspace{-1.5cm}
\hat \Phi_{\hat f} (\hat x, \tilde y, \hat t) =  \nonumber \\
&&\hspace{-1.cm}  
= \int_{(\hat x', \tilde y') \in {\mathbb R} \times {\mathbb R}_+} \hat \xi \big( \frac{\hat \rho^{\tilde \Psi}(\hat x,t)+ \hat \rho^{\tilde \Psi}(\hat x',t)}{2} \big)  \, \frac{1}{\varepsilon} \, \tilde \Psi \big( \frac{|\hat x- \hat x'|}{\varepsilon} \big)
\, \tilde \phi(\tilde y - \tilde y') \, \hat f( \hat x', \tilde y', \hat t) \, d \hat x' \, d \tilde y',
\label{eq:mfl_pot_tilde_finance_2} \\
&&\hspace{-1.5cm}
\hat \rho^{\tilde \Psi}(\hat x, \hat t) =\int_{(\hat x',\tilde y') \in {\mathbb R} \times {\mathbb R}_+} \frac{1}{\varepsilon} \tilde \Psi \big (\frac{|\hat x- \hat x'|}{\varepsilon} \big) \, \, \hat f(\hat x', \tilde y', \hat t) \, d \hat x' \, d \tilde y'. 
\label{eq:mfl_rho_Psi_hat}
\end{eqnarray}
Inserting these changes of variables and unknowns into (\ref{eq:mfl_eq_tilde_finance}) and using (\ref{eq:mfl_force_tilde_finance_2}), we finally obtain the following perturbation problem:
\begin{eqnarray}
&&\hspace{-1cm}
\varepsilon \big( \partial_{\hat t} \hat f + \partial_{\hat x} (\hat V(\hat x, \tilde y) \hat f) \big) + \partial_{\tilde y} (\hat F_{\hat f}\, \hat f) = \tilde d \, \partial^2_{\tilde y} \big( \tilde y^2  \hat f \big).
\label{eq:mfl_eq_eps_finance}
\end{eqnarray}

Now, by Taylor expansion in (\ref{eq:mfl_pot_tilde_finance_2}) and (\ref{eq:mfl_rho_Psi_hat}), and using the normalization condition (\ref{eq:dd_normal}), we get, assuming that $\hat f$ varies only at the large scale: 
\begin{eqnarray}
&&\hspace{-1cm}
\hat \Phi_{\hat f} (\hat x,\tilde y,\hat t) = \xi(\hat \rho(\hat x,\hat t)) \,  \,
\int_{\tilde y' \in {\mathbb R}_+} \, \tilde \phi( \tilde y- \tilde y') \, \hat f ( \hat x, \tilde y', \hat t) \, d  \tilde y'
 + {\mathcal O}(\varepsilon^2).
\label{eq:Ffexpan_finance_0}
\end{eqnarray}
Here, we introduce the local density $\hat \rho(\hat x,\hat t)$ of agents having economic configuration $\hat x$ at time $\hat t$,  $\nu_{\hat x,\hat t}(\tilde y)$ the conditional probability derived from the probability density $\hat f(\hat t)$ by conditioning on the economic configuration variable being equal to $\hat x$, by:
\begin{eqnarray}
&&\hspace{-1cm}
\nu_{\hat x,\hat t}(\tilde y) = \frac{1}{\hat \rho(\hat x,\hat t)} \hat f(\hat x,\tilde y,\hat t), \qquad \hat \rho(\hat x,\hat t) = \int_{\tilde y \in {\mathbb R}_+} f(\hat x,\tilde y,\hat t) \, d \tilde y,
\label{eq:mfl_mom_eps_finance} 
\end{eqnarray}
We assume that $\nu_{\hat x,\hat t} \in {\mathcal P}_{\mbox{\scriptsize ac}}({\mathbb R}_+)$, the space of absolutely continuous probability measures on ${\mathbb R}_+$. With these notations, (\ref{eq:Ffexpan_finance_0}) can be written:
\begin{eqnarray}
&&\hspace{-1cm}
\hat \Phi_{\hat f} (\hat x,\tilde y,\hat t)  = \bar \Phi_{\hat \rho(\hat x,\hat t),\nu_{\hat x,\hat t}}(\tilde y) +  {\mathcal O}(\varepsilon^2),
\label{eq:Ffexpan_finance_00}
\end{eqnarray}
where the functional $\bar \Phi$: $(\hat \rho, \nu) \in {\mathbb R}_+ \times {\mathcal P}_{\mbox{\scriptsize ac}}({\mathbb R}_+) \to \bar \Phi_{\hat \rho, \nu} \in C^2({\mathbb R}_+)$ is defined as follows:
\begin{eqnarray}
&&\hspace{-1cm}
\bar \Phi_{\hat \rho,\nu}(\tilde y) =  \, \hat \rho  \, \hat \xi(\hat \rho) \, \int_{\tilde y' \in {\mathbb R}_+} \, \tilde \phi(\tilde y- \tilde y') \, \nu ( \tilde y') \, d  \tilde y' ,
\label{eq:Ffexpan_finance}
\end{eqnarray}

Eqs. (\ref{eq:Ffexpan_finance_00}), (\ref{eq:Ffexpan_finance}) state that, up to factors of order ${\mathcal O}(\varepsilon^2)$, the cost function is a functional of the conditional probability $\nu_{\hat x,\hat t}$ and of the density $\hat \rho(\hat x,\hat t)$ only, and therefore, only depends on local quantities in the economic configuration variable $\hat x$. The  ${\mathcal O}(\varepsilon^2)$ term collects all non-local effects in economic configuration. These effects are supposed to be much smaller than the local ones. This is an expression of the scale separation in economic configuration: local effects in the economic configuration variable are supposed to have a much bigger influence than non-local ones on a given agent. We comment the relevance of this assumption in the three examples discussed in the introduction. 

In the first example, concerning the influence of migration on the wealth distribution, local trading may be viewed as much more important than long-distance ones, especially when services are concerned. Also, the time-scale of trading is much faster than the time-scale of demographic changes due to migration. Therefore, the time-scale separation is also justified. In the second and third example pertaining with the coupled evolution of wealth and social status on the one hand and wealth and education level on the other hand, similar observations can be made. Wealth exchanges are more likely to be important with a given social status or education level than outside them. A similar observation can be made for the time-scale of trading which is much faster than the time-scales of social status or education level changes. Of course, it is difficult to quantify the amount of `locality' of the wealth exchanges within a social network. Precisely, the present model can actually help testing such hypotheses by providing a coarse-grained description of the system which could be more easily compared to actual data.

Keeping only the leading order term of (\ref{eq:Ffexpan_finance_00}) inside (\ref{eq:mfl_eq_eps_finance}), (\ref{eq:mfl_force_tilde_finance_2}), we are finally led to the following perturbation problem (where we drop all 'hats',  'tildes' and 'bars'):
\begin{eqnarray}
&&\hspace{-1cm}
\varepsilon \big( \partial_t f^\varepsilon + \partial_x (V(x,y) f^\varepsilon) \big) = - \partial_y (F_{f^\varepsilon} \, f^\varepsilon) + d \, \partial^2_y \big( y^2 f^\varepsilon \big),
\label{eq:mfl_eq_eps_finance_2} \\
&&\hspace{-1cm}
F_f (x,y,t) = - \partial_y \Phi_{\rho(x,t), \nu_{x,t}}   \,  ,
\label{eq:Ffexpan_finance_2} \\
&&\hspace{-1cm}
\Phi_{\rho, \nu}  (y) =  \rho  \, \xi(\rho) \, \int_{y' \in {\mathbb R}_+} \, \phi(y-y') \, \nu ( y') \, d  y' ,
\label{eq:Phifexpan_finance_2} \\
&&\hspace{-1cm}
\nu_{x,t}(y) = \frac{1}{\rho(x,t)} f(x,y,t),
\label{eq:nu_def} \\
&&\hspace{-1cm}
\rho(x,t) = \int_{y \in {\mathbb R}_+} f(x,y,t) \, dy.
\label{eq:mfl_mom_eps_finance_2}
\end{eqnarray}
The goal is to derive the limit $\varepsilon \to 0$ of this problem. For this purpose, we first consider the homogeneous economic configuration case in the section below.

%%%%%%%%%%%%%%%%%%%%%%%%%%%%%%%%%%%%%
%%%%%%%%%%%%%%%%%%%%%%%%%%%%%%%%%%%%%
%%%%%%%%%%%%%%%%%%%%%%%%%%%%%%%%%%%%%
%%%%%%%%%%%%%%%%%%%%%%%%%%%%%%%%%%%%%
%%%%%%%%%%%%%%%%%%%%%%%%%%%%%%%%%%%%%
\setcounter{equation}{0}
\section{The homogeneous configuration case}
\label{sec:homogeneous}

%%%%%%%%%%%%%%%%%%%%%%%%%%%%%%%%%%%%%
%%%%%%%%%%%%%%%%%%%%%%%%%%%%%%%%%%%%%
\subsection{The collision operator $Q$}
\label{subsec:homo_general}

Here, we assume that the wealth dynamics is independent of the position in economic configuration space and we restrict the system to the wealth variable $y$ only. Then, $f$ becomes a mapping from $t \in [0,\infty[$ to 
$\ f(t) \in L^1({\mathbb R}_+)$. Obviously, the density 
\begin{eqnarray}
&&\hspace{-1cm}
\rho = \int_{y \in {\mathbb R}_+} f(y,t) \, dy   , 
\label{eq:normaliz_homo}
\end{eqnarray}
is conserved, i.e. $\rho$ is independent of $t$. Consequently, the trading frequency $\rho \xi(\rho)$ is a constant which will be denoted by $\kappa$.  Dividing Eq. (\ref{eq:mfl_eq_eps_finance_2}) by the constant $\rho$, we can write it as an equation for the probability density $\nu = \nu_t = \frac{f(\cdot,t)}{\rho}$ as follows:
\begin{eqnarray}
&&\hspace{-1cm}
\partial_t \nu = Q(\nu), 
\label{eq:mfl_eq_homo_0} \\
&&\hspace{-1cm}
Q(\nu) = - \partial_y (F_{\nu} \, \nu) + d \, \partial_y^2 \big( y^2 \nu \big),
\label{eq:mfl_eq_homo}
\end{eqnarray}
where $F_{\nu}$ is given by
\begin{eqnarray}
&&\hspace{-1cm}
F_{\nu}(y) = - \partial_y \Phi_{\nu} (y),
\label{eq:mfl_force_homo} \\
&&\hspace{-1cm}
\Phi_\nu (y)   = \kappa \int_{y' \in {\mathbb R}_+} \, \phi(y-y') \, \nu ( y') \, d  y' ,
\label{eq:mfl_pot_homo}
\end{eqnarray}
and with initial condition given by $\nu_0$. We have omitted the dependence of $\Phi$ upon $\rho$, since now $\rho$ is a constant. By analogy with classical kinetic theory of gases, we refer to $Q$ as the `collision operator'.  

We now introduce different expressions and a functional setting for $Q$. We start with some definitions.

\begin{definition}
We define $\Xi_\nu$ and $\mu_\nu$: $\nu \in {\mathcal P}_{\mbox{\scriptsize ac}}({\mathbb R}_+) \mapsto \Xi_\nu, \, \mu_\nu \in C^2({\mathbb R}_+)$ such that:
\begin{eqnarray}
&&\hspace{-1cm}
\partial_y \Xi_{\nu}(y) = \frac{1}{y^2} \partial_y \Phi_{\nu}(y) +  \frac{2d}{y} ,
\label{eq:Phi_def_finance_2} \\
&&\hspace{-1cm}
\mu_{\nu}(y) = \Xi_{\nu}(y) + d \ln \nu,
\label{eq:mu_def_finance_3}
\end{eqnarray}
The function $\Xi_\nu$ will be referred to as the 'twisted cost function' and takes into account the geometric effect of the Brownian noise. The function $\mu_\nu(y)$ will be referred to as the 'augmented cost function'. It takes the previous twisted cost function and adds the contribution of the diffusion.
\label{def:Phimu}
\end{definition}

\noindent
We now introduce the Gibbs equilibria, which will appear as the equilibria of $Q$. We note that the importance of the Gibbs equilibria in economics and finance is reviewed in \cite{Yakovenko_Rosser_RevModPhys09}. 

\begin{definition}
For any given cost function $\Xi(y)$, we introduce the 'Gibbs measure' $M_\Xi(y)$ associated to $\Xi$ by:
\begin{eqnarray}
M_\Xi(y) &=& \frac{1}{Z_\Xi} \exp \big( - \frac{\Xi(y)}{d} \big), \qquad Z_\Xi = \int_{y \in {{\mathbb R}_+}} \exp \big( - \frac{\Xi(y)}{d} \big) \, dy.
\label{eq:Von-Mises}
\end{eqnarray}
\label{def:MPhi}
\end{definition}

\begin{remark}
(i) The definition of $Z_{\Xi}$ is such that $M_\Xi$ is a probability density, i.e. it satisfies
\begin{eqnarray}
\int_{y \in {{\mathbb R}_+}} M_\Xi(y) \, dy = 1.
\label{eq:Von-Mises_norm}
\end{eqnarray}

\label{rem:chi_square}
\end{remark}

\noindent
Now, using the definitions above, we can recast $Q$ into various forms below. First, introduce the following definition. 

\begin{definition}
To any twisted cost function $\Xi \in C^2({\mathbb R}_+)$, we associate the linear collision operator ${\mathcal Q}_{\Xi}$ defined by the following equivalent expressions: 
\begin{eqnarray}
{\mathcal Q}_{\Xi}(\nu) (y) &=& \partial_y \Big( y^2  \, \big( \, \partial_y \Xi(y) \, \nu(y) + d \partial_y  \nu(y) \, \big) \, \Big) ,
\label{eq:Q_def_finance_3_0} \\
&=& \partial_y \big( y^2  \, \partial_y \mu(y) \, \nu(y) \big).
\label{eq:Q_def_finance_3_1} \\
&=& d \, \partial_y \Big( y^2 \, M_{\Xi}(y) \, \partial_y \big( \frac{\nu(y) }{M_{\Xi}(y)} \big) \Big),
\label{eq:Q_def_finance_3_3}
\end{eqnarray}
with $\mu = \Xi + d \ln \nu$.
\label{def:mathcalQ}
\end{definition}

\noindent
Then, we have the following lemma, the proof of which is obvious:

\begin{lemma}
The operator $Q$ given by (\ref{eq:mfl_eq_homo}) has the following expressions:
\begin{eqnarray}
Q(\nu) (y) &=& {\mathcal Q}_{\Xi_\nu}(\nu). 
\label{eq:Q_def_finance_3_2}
\end{eqnarray}
where the twisted cost function $\Xi_\nu$ is defined by (\ref{eq:Phi_def_finance_2}). 
\label{lem:QUps_expression}
\end{lemma}

In the next section, we study the properties of $Q$ in more detail. More specifically, we are interested in the equilibria (i.e. the solutions of $Q(\nu) = 0$) and the collision invariants (i.e. the functions $\psi(y)$ which cancel $Q(f)$ upon integration with respect to $y$). These two concepts are key to the derivation of the macroscopic equations.

%%%%%%%%%%%%%%%%%%%%%%%%%%%%%%%%%%%%%
%%%%%%%%%%%%%%%%%%%%%%%%%%%%%%%%%%%%%
\subsection{Properties of $Q$}
\label{subsec:homo_prop_Q}

We introduce a weak form of the operator $Q$. For this purpose, we define the following functional spaces. Let $\Xi \in C^2({\mathbb R}_+)$ be a given twisted cost function. Define
\begin{eqnarray*}
&&\hspace{-1cm}
{\mathcal X}_{\Xi} = \Big\{ u \, \,  \mbox{ such that } \, \,  \int_{y \in {\mathbb R}_+} |u (y)|^2  \,  M_\Xi(y) \, \frac{dy}{y^2} < \infty \Big\}, \\
&&\hspace{-1cm}
{\mathcal H}_{\Xi} = \Big\{ u \, \,  \mbox{ such that } \, \,  \int_{y \in {\mathbb R}_+} \big( |u (y)|^2 + |y^2 \, \partial_y u (y)|^2 \big)  \,  M_\Xi(y) \, \frac{dy}{y^2} < \infty \Big\},
\end{eqnarray*}
endowed with the associated norms $\|u\|_{{\mathcal X}_{\Xi}}$, and  $\|u\|_{{\mathcal H}_{\Xi}}$. We also define
$$ {\mathcal H}_{\Xi 0} = \{ u \in {\mathcal H}_{\Xi} \, \, \mbox{such that } \, \,  \int_{y \in {\mathbb R}_+} u (y) \,  M_\Xi(y) \, \frac{dy}{y^2}= 0 \}. $$
The spaces ${\mathcal X}_{\Xi}$ and ${\mathcal H}_{\Xi}$ are respectively $L^2$ and $H^1$ type spaces associated to the measure $M_\Xi(y) \, \frac{dy}{y^2}$ and the space ${\mathcal H}_{\Xi 0}$ is similar to an $H^1$ space of functions having zero mean with respect to this measure. Therefore, we can expect that a Poincar\'e inequality holds, provided suitable assumptions on $\Xi$ are made. 

At this point, we will assume that such a Poincar\'e inequality holds. In section \ref{subsec:homo_quadra} below, we will prove this Poincar\'e inequality for quadratic trading interactions (the setting of \cite{Bouchaud_Mezard_PhypsicaA00}). 

\begin{assumption}
We assume that the function $\Xi$: $y \in {\mathbb R}_+ \mapsto \Xi(y) \in {\mathbb R}$ is such that $\Xi \in C^2({\mathbb R}_+)$ and that there exists a constant $C>0$ with
\begin{eqnarray}
&&\hspace{-1cm}
|u|_{{\mathcal H}_\Xi} =  \int_{y \in {\mathbb R}_+} |y \, \partial_y u (y)|^2 \,  M_\Xi(y) \, dy \geq C \|u\|_{{\mathcal H}_\Xi}, \quad \forall u \in {\mathcal H}_{\Xi 0}.
\label{eq:poincare}
\end{eqnarray}
\label{ass:poincare}
\end{assumption}

\noindent
This Poincar\'e inequality equivalently states that the semi-norm $|u|_{{\mathcal H}_\Xi}$ is a norm on ${\mathcal H}_{\Xi 0}$ which is equivalent to the norm $\|u\|_{{\mathcal H}_\Xi}$. The following lemma gives some properties of the linear operator ${\mathcal Q}_{\Xi}$. 

\begin{lemma} We suppose that $\Xi \in C^2({\mathbb R}_+)$ is given and satisfies Assumption \ref{ass:poincare}. \\
(i) The weak form of the problem:
\begin{eqnarray}
&&\hspace{-1cm}
``\mbox{Let} \quad g \quad \mbox{be given;} \quad \mbox{find} \quad f \quad \mbox{such that:} \quad {\mathcal Q}_{\Xi} (f)  = g",
\label{eq:problem}
\end{eqnarray}
where $f$ and $g$ are sufficiently smooth functions on ${\mathbb R}_+$ is given by:
\begin{eqnarray}
&&\hspace{-1cm}
``\mbox{Let} \quad \psi \in {\mathcal X}_\Xi \quad \mbox{be given;} \quad
\mbox{find} \quad \varphi \in {\mathcal H}_\Xi \quad \mbox{such that:} \nonumber  \\
&&\hspace{-1.5cm}
\int_{y \in {\mathbb R}_+} \partial_y \varphi (y) \, \partial_y \sigma (y) \, y^2 \,  M_\Xi(y) \, dy =
- \int_{y \in {\mathbb R}_+} \psi \, \sigma(y) \,  M_\Xi(y) \, dy
, \quad \forall \sigma \in {\mathcal H}_\Xi",
\label{eq:GCI_finance_weak_2}
\end{eqnarray}
where
\begin{eqnarray}
&&\hspace{-1cm}
f = M_\Xi \, \varphi, \qquad g = M_\Xi \, \psi.
\label{eq:chgvar}
\end{eqnarray}

\noindent (ii) If $\psi = 0$, the solution space for this problem is spanned by the constants. In terms of $f$, the solution space is the one-dimensional linear space spanned by $M_\Xi$.

\noindent (iii) If $\psi \not= 0$, problem (\ref{eq:GCI_finance_weak_2}) admits a solution if and only if
\begin{eqnarray}
&&\hspace{-1cm}
\int_{y \in {\mathbb R}_+} \psi (y) \,  M_\Xi(y) \, dy= 0 .
\label{eq:solv_cnd}
\end{eqnarray}
If (\ref{eq:solv_cnd}) is satisfied, there exists a unique solution $\varphi_0 \in {\mathcal H}_{\Xi 0}$ and the solution space is the one-dimensional affine space of functions of the form $\varphi_0 + \mbox{Constant}$. In terms of $g$, the solvability condition (\ref{eq:solv_cnd}) is written
\begin{eqnarray}
&&\hspace{-1cm}
\int_{y \in {\mathbb R}_+} g (y) \, dy= 0 ,
\label{eq:solv_cnd_2}
\end{eqnarray}
and the solution space is the one-dimensional affine space of functions of the form $(\varphi_0 + C) M_\Xi$ where $C \in {\mathbb R}$ is an arbitrary constant. 
\label{lem:dual}
\end{lemma}

\medskip
\noindent
The proof is deferred to the appendix \ref{app:proof:dual}

\medskip
Now, for a given twisted cost function $\Xi$,  we investigate the equilibria of the {\bf linear} operator ${\mathcal Q}_\Xi$ defined at~(\ref{eq:Q_def_finance_3_2}).

\begin{definition}
The equilibria of ${\mathcal Q}_\Xi$ are the weak solutions $\nu \in {\mathcal P}_{\mbox{\scriptsize ac}}({\mathbb R}_+)$ of
\begin{equation}
{\mathcal Q}_\Xi(\nu) = 0.
\label{eq:Qrho=0}
\end{equation}
i.e. are the functions $\nu \in {\mathcal P}_{\mbox{\scriptsize ac}}({\mathbb R}_+)$ such that $\theta = \frac{\nu}{M_\Xi}$ belongs to ${\mathcal H}_\Xi$ and satisfies
\begin{eqnarray}
&&\hspace{-1cm}
\int_{y \in {\mathbb R}_+} \partial_y \theta (y) \, \partial_y \sigma (y) \, y^2 \,  M_\Xi(y) \, dy =0, \quad \forall \sigma \in {\mathcal H}_\Xi.
\label{eq:NE_finance_weak}
\end{eqnarray}
\label{def:equi}
\end{definition}

\noindent
We deduce the following:

\begin{proposition}
We suppose that $\Xi$ is given and satisfies Assumption \ref{ass:poincare}. Then $\nu$ is an equilibrium of ${\mathcal Q}_\Xi$ if and only if  $\nu = M_\Xi$ where $M_\Xi$ is given by (\ref{eq:Von-Mises}).
\label{prop:equi_finance}
\end{proposition}

\medskip
\noindent
{\bf Proof.} Let $\nu$ be an equilibrium of ${\mathcal Q}_\Xi$ or equivalently let $\theta = \frac{\nu}{M_\Xi} \in {\mathcal H}_\Xi$ be a weak solution of (\ref{eq:NE_finance_weak}). By Lemma \ref{lem:dual} (ii), the unique solutions are $\theta = \mbox{Constant}$. Hence, by the normalization condition (\ref{eq:normaliz_homo}), the unique weak solution $\nu$ of (\ref{eq:Qrho=0}) in ${\mathcal P}_{\mbox{\scriptsize ac}}({\mathbb R}_+)$ is~$f = M_\Xi$. \endproof

\medskip
We now turn towards defining the equilibria of the nonlinear operator $Q(\nu) = {\mathcal Q}_{\Xi_\nu}(\nu)$, where $\Xi_\nu$ is related to $\nu$ through (\ref{eq:Phi_def_finance_2}). We define:

\begin{definition}
The equilibria of $Q$ are the weak solutions $\nu \in {\mathcal P}_{\mbox{\scriptsize ac}}({\mathbb R}_+)$ of
\begin{equation}
Q(\nu) = {\mathcal Q}_{\Xi_\nu}(\nu) = 0,
\label{eq:Qrho=0_2}
\end{equation}
where $\Xi_\nu$ is related to $\nu$ through (\ref{eq:Phi_def_finance_2}). These equilibria are the functions $\nu \in {\mathcal P}_{\mbox{\scriptsize ac}}({\mathbb R}_+)$ such that $\theta = \frac{\nu}{M_{\Xi_\nu}}$ belongs to ${\mathcal H}_{\Xi_\nu}$ and satisfies
\begin{eqnarray}
&&\hspace{-1cm}
\int_{y \in {\mathbb R}_+} \partial_y \theta (y) \, \partial_y \sigma (y) \, y^2 \,  M_{\Xi_\nu}(y) \, dy =0, \quad \forall \sigma \in {\mathcal H}_{\Xi_\nu}.
\label{eq:NE_finance_weak_2}
\end{eqnarray}
\label{def:equilibria_2}
\end{definition}

\noindent
The equilibria of $Q$ are given in the next:

\begin{proposition}
$\nu$ is an equilibrium of $Q$ if and only if $\nu$ is a solution of the fixed point problem
\begin{eqnarray}
&&\hspace{-1cm}
\nu = M_{\Xi_\nu}. 
\label{eq:NE_fixedpoint}
\end{eqnarray}
\label{prop:equilibria_2}
\end{proposition}

\noindent
{\bf Proof.} If $\nu$ is an equilibrium of $Q$, it is an equilibrium of ${\mathcal Q}_{\Xi}$ for a certain twisted cost function $\Xi$, namely $\Xi = \Xi_\nu$. Then, from Prop. \ref{prop:equi_finance}, it is of the form $\nu = M_{\Xi_\nu}$. Reciprocally, if $\nu$ satisfies (\ref{eq:NE_fixedpoint}), $\nu$ is an equilibrium of ${\mathcal Q}_{\Xi_\nu}$ and so, satisfies ${\mathcal Q}_{\Xi_\nu}(\nu) = 0$. But, because of (\ref{eq:Q_def_finance_3_2}), we have $Q(\nu) = 0$ and $\nu$ is therefore an equilibrium. \endproof

\begin{remark}
In statistical physics, the equilibria are called the Thermodynamic Equilibria. 
\label{rem:statphys}
\end{remark}

The fixed point equation (\ref{eq:NE_fixedpoint}) can be equivalently stated as a fixed point equation for the twisted cost function $\Xi$. This equivalence is established in the following proposition, the proof of which is obvious. 

\begin{proposition}
$\nu$ is an equilibrium of $Q$ if and only if $\nu = M_\Xi$ where the twisted cost function $\Xi$ is a solution of the fixed point problem:
\begin{eqnarray}
&&\hspace{-1cm}
\Xi = \Xi_{M_\Xi}. 
\label{eq:NE_fixedpoint_2}
\end{eqnarray}
\label{prop:equilibria_3}
\end{proposition}

Prop. \ref{prop:equilibria_2} or \ref{prop:equilibria_3} do not say anything about the existence and uniqueness of the equilibria. Such information requires the analysis of the fixed point equations (\ref{eq:NE_fixedpoint}) or (\ref{eq:NE_fixedpoint_2}), and necessitates specific assumptions on the trading interaction potential $\phi$. This analysis is outside the scope of the present work. We now provide a game theoretical interpretation of the equilibria in the next section.

%%%%%%%%%%%%%%%%%%%%%%%%%%%%%%%%%%%%%
%%%%%%%%%%%%%%%%%%%%%%%%%%%%%%%%%%%%%
\subsection{Game theoretical interpretation}
\label{subsec:homo_Nash}

In this section, we show that the equilibria $M_\Xi$ are Nash equilibria \cite{Nash_PNAS50} for the mean-field game (also known as non-cooperative non-atomic anonymous game with a continuum of players \cite{Cardaliaguet_NotesCollegeFrance12}) associated to the cost function $\mu_\nu(y)$ given by (\ref{eq:mu_def_finance_3}). For such a game, a Nash equilibrium measure $\nu_{\mbox{\scriptsize{NE}}} \in {\mathcal P}({\mathbb R}_+)$ is such that \cite{Cardaliaguet_NotesCollegeFrance12} (see also \cite{Blanchet_Carlier, Blanchet_Mossay_Santambrogio}) there exists a constant $K$ and
\begin{equation} \left\{ \begin{array}{ll} \mu_{\nu_{\mbox{\tiny{NE}}}}(y) = K & \qquad \forall y \in \mbox{Supp} (\nu_{\mbox{\tiny{NE}}}), \\
\mu_{\nu_{\mbox{\tiny{NE}}}}(y) \geq K & \qquad \forall y \in {\mathbb R}_+\, , \end{array} \right.
\label{eq:NE}
\end{equation}
where Supp$(\nu_{\mbox{\tiny{NE}}})$ denotes the support of $\nu_{\mbox{\tiny{NE}}}$. This definition is equivalent to the following statement \cite{Cardaliaguet_NotesCollegeFrance12}:
\begin{equation}
\int_{y \in {\mathbb R}_+} \mu_{\nu_{\mbox{\tiny{NE}}}}(y) \, \nu_{\mbox{\tiny{NE}}}(y) \, dy = \inf_{\nu \in {\mathcal P}_{\mbox{\scriptsize{ac}}}({\mathbb R}_+)} \int_{y \in {\mathbb R}_+} \mu_{\nu_{\mbox{\tiny{NE}}}}(y) \, \nu(y) \, dy .
\label{eq:NE2}
\end{equation}
Eq. (\ref{eq:NE2}) is called the 'mean-field' equation. Now, we have the following

\medskip

\begin{theorem}
Let $\nu \in {\mathcal P}_{\mbox{\scriptsize ac}}({\mathbb R}_+)$. Then the two following statements are equivalent:\\
\indent (i) $\nu$ is an equilibrium (\ref{eq:NE_fixedpoint}), \\
\indent (ii) $\nu$ is a Nash equilibrium (\ref{eq:NE}) for the Mean-Field game associated to the cost function $\mu_\nu$ given by (\ref{eq:mu_def_finance_3}). 
\label{thm:NE}
\end{theorem}

\medskip
\noindent
{\bf Proof.} The proof is identical to that of Theorem 3.5 of \cite{Degond_etal_preprint13_2}. We reproduce it here for the sake of completeness. 

\medskip
\noindent
(i) $\Rightarrow$ (ii). Let $\nu=M_{\Xi_\nu}$ be an equilibrium (\ref{eq:NE_fixedpoint}). Then, we have by (\ref{eq:mu_def_finance_3}):
\begin{eqnarray*}
\mu_\nu(y) = - d \, \ln Z_{\Xi_\nu} = \mbox{ Constant, } \quad \forall y \in {\mathbb R}_+ .
\end{eqnarray*}
and hence, $\nu$ satisfies condition (\ref{eq:NE}) characterizing Nash equilibria.

\medskip
\noindent
(ii) $\Rightarrow$ (i). Let $\nu_{\mbox{\tiny{NE}}}$ be a Nash equilibrium (\ref{eq:NE}). We show that $\mbox{Supp} (\nu_{\mbox{\tiny{NE}}}) = {\mathbb R}_+$. Indeed, by contradiction, suppose $\mbox{Supp} (\nu_{\mbox{\tiny{NE}}}) \varsubsetneq {\mathbb R}_+$. There exists $y \in {\mathbb R}_+$ such that $\nu_{\mbox{\tiny{NE}}}(y) = 0$. Then, because of the log inside (\ref{eq:mu_def_finance_3}) and the finiteness of $\Xi_{\nu_{\mbox{\tiny{NE}}}}(y)$ for $0<y<\infty$, we have $\mu_{\nu_{\mbox{\tiny{NE}}}} (y) = - \infty$ which is a contradiction to the second line of (\ref{eq:NE}). Therefore, by the first line of (\ref{eq:NE}),  $\mu_{\nu_{\mbox{\tiny{NE}}}}$ is identically constant over the entire space ${\mathbb R}_+$. From the expression of $\mu_{\nu_{\mbox{\tiny{NE}}}}$ in (\ref{eq:mu_def_finance_3}), $ \nu_{\mbox{\tiny{NE}}}$ is proportional to $\exp (- \Xi_{\nu_{\mbox{\tiny{NE}}}}(y)/d )$, which means that it is an equilibrium (\ref{eq:NE_fixedpoint}). \endproof

The mean-field model (\ref{eq:mfl_eq_homo})-(\ref{eq:mfl_pot_homo}) can be recast as a transport equation as follows
\begin{eqnarray}
&&\hspace{-1cm}
\partial_t \nu + \partial_y \cdot (v \, \nu) = 0,
\label{eq:mfl_transport} \\
&&\hspace{-1cm}
v = - \partial_y \mu_\nu.
\label{eq:mfl_drift}
\end{eqnarray}
It describes the bulk motion of agents which move in the direction of the steepest descent towards the minimum of $\mu_\nu$. When all agents have reached the minimum of $\mu_\nu$, then $\mu_\nu$ is a constant and describes a Nash Equilibrium. Therefore, in the proposed dynamics, the agents' actions result in a motion in the steepest descent direction towards the Nash equilibrium.

Let us now interpret the fixed point problems (\ref{eq:NE_fixedpoint}) or (\ref{eq:NE_fixedpoint_2}) in the framework of this Nash equilibrium problem. Let first the agents be distributed according to the distribution $\nu$. Each of the them constructs a strategy by forming the twisted cost function $\Xi_\nu$ calculated from $\nu$ through (\ref{eq:Phi_def_finance_2}). This cost function enables each agent to enforce a best-reply strategy and move downwards the gradient of $\mu_\nu$ until reaching its minimum. By the discussion above, when all the agents have reached the minimum of $\mu_\nu$, their distribution is given by $M_{\Xi_\nu}$. Now, this can only be an equilibrium if, when constructing a new twisted cost function $\Xi_{M_{\Xi_\nu}}$ based on this new distribution, we find the same twisted cost function as before, i.e. $\Xi_\nu$. Indeed, otherwise, the new cost function will trigger a new motion along the gradients of the augmented cost function $\mu_{M_{\Xi_\nu}}$ which will destroy the equilibrium $M_{\Xi_\nu}$. Therefore, a condition for $M_{\Xi}$ to be an equilibrium is that $\Xi_{M_\Xi} = \Xi$ which is (\ref{eq:NE_fixedpoint_2}). Within this condition, no agent can find a better strategy, since when forming the twisted cost function associated to this distribution, he finds the same cost as before which is already minimized.

The important conclusion of this discussion is that the fixed point problems (\ref{eq:NE_fixedpoint}) or (\ref{eq:NE_fixedpoint_2}) express that the agents have found a Nash equilibrium associated to the augmented cost function $\mu_\nu$.

\begin{remark}
In \cite{Degond_etal_preprint13_2}, a variational structure for potential games \cite{Monderer_Shapley_GamesEconomicBehav96}. is developed. A variational structure supposes that there exists a functional ${\mathcal U} (\nu)$ such that
$$ \Xi_\nu(y) = \frac{\delta {\mathcal U} (\nu)}{\delta \nu}(y), \quad \forall y \in {\mathbb R}_+, $$
where $\frac{\delta {\mathcal U} (\nu)}{\delta \nu}$ is the functional derivative of ${\mathcal U}$ defined by
\begin{equation*}
\int_{y \in {\mathbb R}_+} \frac{\delta {\mathcal U} (\nu)}{\delta \nu}(y) \, (\varphi(y) - \nu(y)) \, dy = \lim_{s \to 0^+} \frac{1}{s} ({\mathcal U} (\nu + s (\varphi-\nu)) - {\mathcal U} (\nu)),
%\label{eq:fct_der}
\end{equation*}
for any test function $\varphi(y) \in {\mathcal P}_{\mbox{\scriptsize ac}}({\mathbb R}_+)$. The existence of such a functional gives a very strong constraint on $\Xi$. In the present case, we have not been able to find such a functional ${\mathcal U} (\nu)$. It seems that the present model provides an example of Mean-Field Game with no potential structure.
\label{rem:variational}
\end{remark}

The existence and uniqueness of Nash equilibria in the most general trading interaction setting will be investigated in future work. We now examine the case of the quadratic trading interaction setting of \cite{Bouchaud_Mezard_PhypsicaA00}.

%%%%%%%%%%%%%%%%%%%%%%%%%%%%%%%%%%%%%
%%%%%%%%%%%%%%%%%%%%%%%%%%%%%%%%%%%%%
\subsection{Quadratic trading interaction}
\label{subsec:homo_quadra}

In this case, we let $\phi(y) = \frac{|y|^2}{2}$. From (\ref{eq:mfl_pot_homo}), we have 
\begin{eqnarray}
&&\hspace{-1cm}
\partial_y \Phi_\nu (y) = \kappa \int_{y' \in {\mathbb R}_+} \, (y-y') \, \nu ( y') \, d  y' = \kappa (y - \Upsilon_\nu),
\label{eq:mf_pot_quadra}
\end{eqnarray}
where $\Upsilon_\nu$ is the mean wealth associated to the wealth distribution $\nu$, given by: 
\begin{eqnarray}
&&\hspace{-1cm}
\Upsilon_\nu = \int_{y \in {\mathbb R}_+} \, y \, \nu ( y) \, d  y  .
\label{eq:mean_wealth}
\end{eqnarray}
We see that the trading cost function $\Phi_\nu$ only depends on the mean wealth, i.e. the first moment of the distribution $\nu$, instead of depending on the whole functional shape of $\nu$ as in the general trading interaction case treated so far. We will exploit this fact and express all objects related to this interaction in terms of $\Upsilon_\nu$ only. On the other hand, this shows that the quadratic trading interaction case is a degenerate case. Conclusions drawn from this case might be non-generic and misleading.

The operator $Q$ can be expressed as follows (which is similar to the Bouchaud \& M\'ezart form \cite{Bouchaud_Mezard_PhypsicaA00}): 
\begin{eqnarray}
&&\hspace{-1cm}
Q(\nu) (y) = {\mathcal Q}_{\Upsilon_\nu} (\nu) (y) ,
\label{eq:def_mathcalQ_BM_1}
\end{eqnarray}
where, for all $\Upsilon \in {\mathbb R}_+$, we denote by ${\mathcal Q}_\Upsilon (\nu)$ the following linear operator: 
\begin{eqnarray}
&&\hspace{-1cm}
{\mathcal Q}_\Upsilon (\nu) (y) = \partial_y \Big( \kappa \, (y -  \Upsilon ) \,  \nu  + d \,  \partial_y (y^2 \nu) \Big) .
\label{eq:def_mathcalQ_BM_2}
\end{eqnarray}
From (\ref{eq:Phi_def_finance_2}), (\ref{eq:mu_def_finance_3}), the twisted and augmented cost functions are given by: 
\begin{eqnarray}
&&\hspace{-1cm}
\Xi_{\nu}(y) = \tilde \Xi_{\Upsilon_\nu}(y), \qquad \mu_{\nu}(y) = \tilde \Xi_{\Upsilon_\nu}(y) + d \ln \nu,
\label{eq:mu_def_finance_BM_1}
\end{eqnarray}
where, for all $\Upsilon \in {\mathbb R}_+$, we denote by
\begin{eqnarray}
&&\hspace{-1cm}
\tilde \Xi_\Upsilon(y) = (\kappa + 2d) \ln y + \kappa \frac{\Upsilon}{y}.
\label{eq:mu_def_finance_BM_2}
\end{eqnarray}
In the remainder, we will omit the tilde on $\tilde \Xi_\Upsilon$ when the context is clear. 

The Gibbs measure $M_\Upsilon$ associated with $\Xi_\Upsilon$ given by (\ref{eq:mu_def_finance_BM_2}) through (\ref{eq:Von-Mises}) is expressed by
\begin{eqnarray}
&&\hspace{-1cm}
M_{\Upsilon} (y) = 
\frac{1}{Z_{\Upsilon}} \, \,  \frac{1}{y^{1+ \frac{\kappa + d}{d}}} \,
\exp \big( - \frac{\kappa \Upsilon}{d y} \big) ,
\label{eq:NE_finance}
\end{eqnarray}
where $Z_{\Upsilon}$ is given by:
$$ Z_{\Upsilon} = \int_0^\infty \frac{1}{y^{1+ \frac{\kappa + d}{d}}} \, \exp \big( - \frac{\kappa \Upsilon}{d\,y} \big) \, dy. $$
It is well-defined for $\kappa+d>0$. We have
\begin{eqnarray*}
Z_{\Upsilon} = \frac{\Gamma \big(\frac{\kappa+d}{d} \big)}{\big( \frac{\kappa \Upsilon}{d} \big)^{\frac{\kappa+d}{d}}},
\end{eqnarray*}
where $\Gamma$ is the Euler Gamma function.

The condition $\kappa + d >0$ guarantees the integrability of $M_\Upsilon$ when $y \to \infty$. The condition $\kappa>0$ implies that $M_\Upsilon(0) = 0$ because of the exponential factor. We have 
$$M_\Upsilon = g_{\frac{\kappa + d}{d}, \frac{\kappa \Upsilon}{d}}, $$ 
where $g_{\alpha, \beta}$  is  the inverse Gamma distribution
\begin{eqnarray*}
&&\hspace{-1cm}
g_{\alpha, \beta} (y) = \frac{\beta^\alpha}{\Gamma(\alpha)} \frac{1}{y^{1+\alpha}} \, e^{- \frac{\beta}{y}},
\end{eqnarray*}
with shape parameter $\alpha$ and scale parameter $\beta$. This distribution is also sometimes called the scaled inverse chi-squared distribution. It is related to the Gamma distribution
\begin{eqnarray}
&&\hspace{-1cm}
\gamma_{\alpha, \beta} (z) = \frac{\beta^\alpha}{\Gamma(\alpha)} z^{\alpha-1} \, e^{- \beta z},
\label{eq:gamma_distrib}
\end{eqnarray}
by the change of variables $z=1/y$, i.e.
\begin{eqnarray}
&&\hspace{-1cm}
g_{\alpha, \beta} (y) \, dy = \gamma_{\alpha, \beta} (z) \, dz.
\label{eq:link_gamma}
\end{eqnarray}
This distribution has been previously found in \cite{Bouchaud_Mezard_PhypsicaA00, Cordier_etal_JSP05, During_Toscani_PhysicaA07}. When $y$ is large, the distribution $M_\Upsilon$ becomes the Pareto power law distribution, which has a very strong agreement with economic data (see e.g. the review in \cite{Yakovenko_Rosser_RevModPhys09}).

We note the important following consistency property, which is not a priori obvious:

\begin{lemma}
Let $\Upsilon \in {\mathbb R}_+$ be given and let $M_\Upsilon$ be the equilibrium (\ref{eq:NE_finance}). Then, the mean wealth of $M_\Upsilon$ exists and is given by:
\begin{equation}
\Upsilon_{M_\Upsilon} = \Upsilon.
\label{eq:mean_wealth_2}
\end{equation}
\label{lem:mean_wealth_equi}
\end{lemma}

\medskip
\noindent
{\bf Proof.} We compute:
\begin{eqnarray}
&&\hspace{-1cm}
\Upsilon_{M_\Upsilon} = \frac{\displaystyle \int_0^\infty \frac{1}{y^{\frac{\kappa + d}{d}}} \, \exp \big( - \frac{\kappa \Upsilon}{d y} \big) \, dy}{\displaystyle  \int_0^\infty \frac{1}{y^{1+ \frac{\kappa + d}{d}}} \, \exp \big( - \frac{\kappa \Upsilon}{d y} \big) \, dy}.
\label{eq:comp_rel_finance}
\end{eqnarray}
The assumption $\kappa >0$ guarantees that the integrals at both the numerator and the denominator of (\ref{eq:comp_rel_finance}) converge. Then, by the change of variables $z = \frac{\kappa \Upsilon}{yd}$ in (\ref{eq:comp_rel_finance}) and integration by parts, it is straightforward to show that (\ref{eq:mean_wealth_2}) holds. \endproof

In this case, Assumption \ref{ass:poincare} is a theorem. More precisely, we introduce the functional setting: 
\begin{eqnarray}
&&\hspace{-1cm}
{\mathcal X}_\Upsilon = \Big\{ u \, \,  \mbox{ such that } \, \,  \int_{y \in {\mathbb R}_+} |u (y)|^2  \,  M_\Upsilon(y) \, \frac{dy}{y^2} < \infty \Big\}, 
\label{eq:def_XU} \\
&&\hspace{-1cm}
{\mathcal H}_\Upsilon = \Big\{ u \, \,  \mbox{ such that } \, \,  \int_{y \in {\mathbb R}_+} \big( |u (y)|^2 + |y^2 \, \partial_y u (y)|^2 \big)  \,  M_\Upsilon(y) \, \frac{dy}{y^2} < \infty \Big\},
\label{eq:def_HU} 
\end{eqnarray}
endowed with the associated norms $\|u\|_{{\mathcal X}_\Upsilon}$, and  $\|u\|_{{\mathcal H}_\Upsilon}$. We also define
\begin{eqnarray}
&&\hspace{-1cm}
{\mathcal H}_{\Upsilon 0} = \{ u \in {\mathcal H}_\Upsilon \, \, \mbox{such that } \, \,  \int_{y \in {\mathbb R}_+} u (y) \,  M_\Upsilon(y) \, \frac{dy}{y^2}= 0 \}. 
\label{eq:def_HU0} 
\end{eqnarray}
Now, we have the following lemma, which directly follows from the corresponding modified Poincaré inequality for the Gamma distribution \cite{Benaim_Rossignol_arxiv}, formula (10) (see also \cite{Benaim_Rossignol_AnnIHPPS08, Ledoux_LN}):

\begin{lemma}
There exists a constant $C>0$ such that
\begin{eqnarray}
&&\hspace{-1cm}
|u|_{{\mathcal H}_\Upsilon} =  \int_{y \in {\mathbb R}_+} |y \, \partial_y u (y)|^2 \,  M_\Upsilon(y) \, dy \geq C \|u\|_{{\mathcal H}_\Upsilon}, \quad \forall u \in {\mathcal H}_{\Upsilon 0}.
\label{eq:poincare_BM}
\end{eqnarray}
\label{lem_poincare}
\end{lemma}

\noindent
The proof is deferred to Appendix \ref{app:proof:poincare}.

\medskip
Then, Lemma \ref{lem:dual} can be applied and we deduce the following lemma which lists all the equilibria of the linear operator ${\mathcal Q}_\Upsilon$:

\begin{proposition}
The distribution $\nu$ is an equilibrium of ${\mathcal Q}_\Upsilon$ if and only if  $\nu = M_\Upsilon$ where $M_\Upsilon$ is given by (\ref{eq:NE_finance}).
\label{prop:equi_finance_BM}
\end{proposition}

We now turn to the equilibria of the nonlinear operator $Q(\nu) = {\mathcal Q}_{\Upsilon_\nu}(\nu)$. They are characterized in the following proposition:

\begin{proposition}
$\nu$ is an equilibrium of $Q$ if and only if there exists $\Upsilon \in {\mathbb R}_+$ such that $\nu = M_{\Upsilon}$. 
\label{prop:equilibria_2_BM}
\end{proposition}

\noindent
{\bf Proof.} If $\nu$ is an equilibrium of $Q$, thanks to (\ref{eq:def_mathcalQ_BM_1}), it is an equilibrium of ${\mathcal Q}_\Upsilon$ for a certain value $\Upsilon$. Theerfore, it is of the form $\nu = M_\Upsilon$ for this value of $\Upsilon$, thanks to Prop. \ref{prop:equi_finance_BM}. Reciprocally, if $\nu = M_\Upsilon$, then, $\nu$ is an equilibrium of ${\mathcal Q}_\Upsilon$ by Prop. \ref{prop:equi_finance_BM} and so, satisfies ${\mathcal Q}_\Upsilon(\nu) = 0$. But, thanks to Lemma \ref{lem:mean_wealth_equi}, we have $\Upsilon_\nu = \Upsilon$. So, $\nu$ is a solution of ${\mathcal Q}_{\Upsilon_\nu}(\nu) = 0$, i.e. a solution of $Q(\nu) = 0$ and is therefore an equilibrium. \endproof

\begin{remark}
In the present case, the fixed point equation (\ref{eq:NE_fixedpoint_2}) reduces to (\ref{eq:mean_wealth_2}). Indeed, since the twisted cost function only depends on the mean wealth $\Upsilon$, Eq. (\ref{eq:NE_fixedpoint_2}) reduces to a fixed point equation on the mean wealth. This equation is deduced from (\ref{eq:NE_fixedpoint_2}) by replacing $\Xi$ by $\Upsilon$, i.e. it takes the form of Eq. (\ref{eq:mean_wealth_2}). But precisely, Lemma \ref{lem:mean_wealth_equi} states that this equation is satisfied for any value of $\Upsilon$. Therefore, for a given value of the mean wealth, there exists a unique Nash equilibrium and the set of Nash equilibria is parametrized by the mean wealth. In the case of more general trading interactions, the situation can be much more complex. Therefore, the quadratic trading interaction case seems somehow degenerate. 
\label{rem:fixedpoint_BM}
\end{remark}

%%%%%%%%%%%%%%%%%%%%%%%%%%%%%%%%%%%%%
%%%%%%%%%%%%%%%%%%%%%%%%%%%%%%%%%%%%%
%%%%%%%%%%%%%%%%%%%%%%%%%%%%%%%%%%%%%
%%%%%%%%%%%%%%%%%%%%%%%%%%%%%%%%%%%%%
%%%%%%%%%%%%%%%%%%%%%%%%%%%%%%%%%%%%%
\setcounter{equation}{0}
\section{The inhomogeneous configuration case: derivation of the macroscopic model}
\label{sec:hydro}

%%%%%%%%%%%%%%%%%%%%%%%%%%%%%%%%%%%%%
%%%%%%%%%%%%%%%%%%%%%%%%%%%%%%%%%%%%%
\subsection{Framework}
\label{subsec_inhomo_setting}

Now, we return to the model in the inhomogeneous configuration case (\ref{eq:mfl_eq_eps_finance_2})-(\ref{eq:mfl_mom_eps_finance_2}) where the positions of the agents in the economic configuration space is considered. The goal of this section is to investigate the limit $\varepsilon \to 0$ of this system. It will enable us to describe the ensemble motion of the agents at large time scales, averaging out over their individual wealth variables. Therefore, we will introduce a suitable coarse-graining procedure. Taking advantage that at large times, individuals relax their wealth variables towards that corresponding to a global Nash equilibrium given by (\ref{eq:NE_fixedpoint}), we use this equilibrium as a prescription for the internal wealth variable distribution of the agents. In this section, we provide the details of this coarse-graining process, known as the hydrodynamic limit in kinetic theory. In a first part, we will consider general trading interaction potentials $\phi$. In the second part, we will focus on the particular case of quadratic trading interactions and will recover a result of \cite{During_Toscani_PhysicaA07}. 

In order to simplify the computations, we make the following assumption, which is not essential but simplifies the computations: 

\begin{assumption}
We assume that the trading frequency $\rho \xi(\rho)$ is a constant given by
\begin{equation}
\rho \xi(\rho) = \kappa.
\label{eq:alpharho=kappa}
\end{equation}
\label{ass:alpharho=constant}
\end{assumption}

Under this assumption, the cost function $\Phi_\nu$ becomes independent of $\rho$ and reduces to (\ref{eq:mfl_pot_homo}), which, in the present spatially inhomogenous setting, is written:
\begin{eqnarray}
&&\hspace{-1cm}
\Phi_{\nu_{x,t}} (y)   = \kappa \int_{y' \in {\mathbb R}_+} \, \phi(y-y') \, \nu_{x,t} ( y') \, d  y' ,
\label{eq:mfl_pot_inhomo}
\end{eqnarray}
We write (\ref{eq:mfl_eq_eps_finance_2}) as
\begin{eqnarray}
&&\hspace{-1cm}
\partial_t f^\varepsilon + \partial_x (V(x,y) f^\varepsilon) = \frac{1}{\varepsilon} Q(f^\varepsilon).
\label{eq:mfl_eq_eps_2_finance}
\end{eqnarray}
The interaction operator $Q$ is given by
\begin{eqnarray}
&&\hspace{-1cm}
Q(f) = {\mathcal Q}_{\Xi_{\nu_{x,t}}} (\nu_{x,\,t}) ,
\label{eq:Q_def_finance_1}
\end{eqnarray}
where, for $\Xi \in C^2({\mathbb R}_+)$ and $\nu \in {\mathcal P}_{\mbox{\scriptsize ac}}({\mathbb R}_+) \cap C^2({\mathbb R}_+)$, ${\mathcal Q}_\Xi (\nu)$ is given by (\ref{eq:Q_def_finance_3_0})-(\ref{eq:Q_def_finance_3_3}), and where $\Xi_{\nu_{x,t}}$ is related to $\Phi_{\nu_{x,t}}$ by (\ref{eq:Phi_def_finance_2}). We recall that $\nu_{x,\,t}$ and $\rho(x,t)$ are given by (\ref{eq:nu_def}), (\ref{eq:mfl_mom_eps_finance_2}).

Here again, we emphasize that $(x,t)$ now refers to slow variables. The left-hand side of (\ref{eq:mfl_eq_eps_2_finance}) describes how the distribution of agents as a function of economic neighborhood and
time evolves. This evolution is driven by the fast, local evolution of this distribution as a function of individual wealth $y$ described by the right-hand side. The parameter $\varepsilon$ at the denominator highlights the fact that the exchanges of wealth occur on a much faster time-scale than the evolution of the agents in the economic configuration variable. The fast evolution of the wealth drives the system towards an equilibrium, i.e., a solution of $Q(f) = 0$. Such a solution is referred to in physics as a Local Thermodynamical Equilibrium (LTE). Below, we use the results of the previous section to show that, in this case, the LTE's are Nash equilibria given in Propositions \ref{prop:equi_finance} and \ref{prop:equilibria_2}.

%%%%%%%%%%%%%%%%%%%%%%%%%%%%%%%%%%%%%
%%%%%%%%%%%%%%%%%%%%%%%%%%%%%%%%%%%%%
\subsection{Local Thermodynamical Equilibria and conservations}
\label{subsec_LTE}

Applying Proposition \ref{prop:equilibria_2} to $Q$ given by (\ref{eq:Q_def_finance_1}), we immediately get the

\begin{corollary}
Let $f:$ $(x,y,t) \in {\mathbb R} \times {\mathbb R}_+ \times {\mathbb R}_+ \mapsto f(x,y,t) \in {\mathbb R}_+$ such that $\forall (x,t) \in {\mathbb R} \times {\mathbb R}_+$, the function $y \mapsto f(x,y,t)$ belongs to $L^1({\mathbb R}_+) \cap C^2({\mathbb R}_+)$. Then, the following two statements are equivalent: \\
(i) $Q(f) = 0$ in the weak sense, where $Q$ is the collision operator (\ref{eq:Q_def_finance_1}), \\
(ii) there exists $\rho(x,t) >0$ and $\Xi_{x,\,t} \in C^2({\mathbb R}_+)$ such that $f$ is given by
\begin{eqnarray}
&&\hspace{-1cm}
f(x,y,t) = F_{\mbox{\scriptsize{eq}}, \, \rho(x,t), \Xi_{x,\,t}} (y) :=\rho(x,t) \, M_{\Xi_{x,\,t}}(y), \quad \forall (x,y,t) \in {\mathbb R} \times {\mathbb R}_+ \times {\mathbb R}_+,
\label{eq:equi_f}
\end{eqnarray}
where, for any $\Xi \in C^2({\mathbb R}_+)$, $M_\Xi$ denotes the Gibbs distribution (\ref{eq:Von-Mises}). 
Additionally, $\Xi_{x,\,t}$ must satisfy the fixed point equation which characterizes the equilibria, namely:
\begin{eqnarray}
&&\hspace{-1cm}
\Xi_{M_{\Xi_{x,\,t}}}=\Xi_{x,\,t}, \quad \forall (x,t) \in {\mathbb R} \times {\mathbb R}_+. 
\label{eq:equi_f_inhomo}
\end{eqnarray}
\label{cor:equi}
\end{corollary}

\begin{remark}
The functions $F_{\mbox{\scriptsize{eq}}, \, \rho, \Xi}$ are the LTE of our wealth distribution dynamical model and they depend on the local density $\rho$ and on the local twisted cost function $\Xi_{x,\,t}$ at given location $x$ in economic configuration and given time $t>0$.
\label{rem:equi}
\end{remark}

We introduce the mean wealth $\Upsilon(x,t)$ associated with $F_{\mbox{\scriptsize{eq}}, \, \rho, \Xi}$, defined by 
\begin{eqnarray}
&& \hspace{-1cm} \Upsilon(x,t) =  \int_{y \in {\mathbb R}_+} y \, M_{\Xi_{x,t}}(y) \, dy.
\label{eq:mw_equi} 
\end{eqnarray}
Now, we can state a first result for the coarse-graining limit $\varepsilon \to 0$ inside Eq. (\ref{eq:mfl_eq_eps_2_finance}). We have the

\begin{theorem}
Suppose that the solution $f^\varepsilon$ to (\ref{eq:mfl_eq_eps_2_finance}) converges to a function $f$ when $\varepsilon \to 0$ smoothly, which means in particular that all derivatives of $f^\varepsilon$ converge to the corresponding derivative of $f$. Then, formally $f$ is given by an LTE (\ref{eq:equi_f}). The agent density $\rho(x,t)$ and wealth density $( \rho \Upsilon )(x,t)$ satisfy the following conservation law:
\begin{eqnarray}
&& \hspace{-1cm} 
\partial_t \rho + \nabla_x \cdot \big(\rho \, u(x;\Xi_{x,t}) \big) = 0 , 
\label{eq:mass_finance} \\
&& \hspace{-1cm} 
\partial_t (\rho \Upsilon) + \nabla_x \cdot \big(\rho \, {\mathcal E}(x;\Xi_{x,t}) \big) = 0 , 
\label{eq:mw_finance} 
\end{eqnarray}
where, for any twisted cost function $\Xi \in C^2({\mathbb R})$, we denote by 
\begin{eqnarray}
&& \hspace{-1cm} 
u (x;\Xi) = \int_{y \in {\mathbb R}_+} V(x,y) \, M_\Xi(y) \, dy,
\label{eq:mean_vel_equi_finance} \\
&& \hspace{-1cm} 
{\mathcal E} (x;\Xi) = \int_{y \in {\mathbb R}_+} y \, V(x,y) \, M_\Xi(y) \, dy.
\label{eq:mw_vel_equi_finance}
\end{eqnarray}
The quantities $u (x;\Xi)$ and ${\mathcal E} (x;\Xi)$ are respectively the agent's and wealth average velocity in configuration space for an equilibrium associated with the twisted cost function $\Xi$. 
\label{thm:limit_epsto0_finance}
\end{theorem}

\medskip
\noindent
{\bf Proof.} From (\ref{eq:mfl_eq_eps_2_finance}), we have that $Q(f^\varepsilon) = {\mathcal O}(\varepsilon)$ and owing to the convergence assumptions made on $f^\varepsilon$, we have $Q(f) = 0$. Thanks to Corollary \ref{cor:equi}, $f$ is of the form (\ref{eq:equi_f}). Now, observe that $1$ and $y$ are collisional invariant of $Q$, meaning that
\begin{eqnarray}
&& \hspace{-1cm} 
\int_{y \in {\mathbb R}_+} Q(f) (y) \, dy = 0, \qquad 
\int_{y \in {\mathbb R}_+} Q(f) (y) \, y \, dy = 0,
\label{eq:inhomo_CI_gene}
\end{eqnarray}
for all functions $f(y)$. The first relation (\ref{eq:inhomo_CI_gene}) easily comes upon integrating (\ref{eq:mfl_eq_homo}) with respect to $y$ and using Green's formula (see also (\ref{eq:solv_cnd_2})). The second relation (\ref{eq:inhomo_CI_gene}) is a consequence of the evenness of the trading interaction potential $\phi$ (see Assumption \ref{ass:even}). Indeed, multiplying (\ref{eq:mfl_eq_homo}) by $y$, integrating it with respect to $y$ and using Green's formula leads to:
\begin{eqnarray}
&& \hspace{-1cm} 
\int_{y \in {\mathbb R}_+} Q(f) (y) \, y \, dy = - \int_{y \in {\mathbb R}_+} \partial_y \Phi_{\nu_{x,t}} \, \nu_{x,t}(y) \, dy \nonumber \\
&& \hspace{1.5cm} 
= - \kappa \int_{(y,y') \in ({\mathbb R}_+)^2} \partial_y \phi(y-y') \, \nu_{x,t}(y) \, \nu_{x,t}(y') \, dy \, dy' = 0, 
\label{eq:inhomo_CI_gene_calcul}
\end{eqnarray}
where we have used the fact that $\partial_y \phi$ is odd and we recall that $\nu_{x,t}$ is related to $f$ by~(\ref{eq:nu_def}). 

Therefore, integrating (\ref{eq:mfl_eq_eps_2_finance}) with respect to $y$ leads to
\begin{eqnarray}
&& \hspace{-1cm} 
\partial_t \rho^\varepsilon + \nabla_x \cdot (\rho^\varepsilon u^\varepsilon) = 0 , \label{eq:mass_eps}
\end{eqnarray}
where $\rho^\varepsilon$ and $\nu^\varepsilon_{x,t}(y)$ are obtained from $f^\varepsilon$ through (\ref{eq:mfl_mom_eps_finance_2}) and (\ref{eq:nu_def}) respectively, and~$u^\varepsilon$, the mean velocity in configuration space associated to $\nu^\varepsilon_{x,t}$, is defined by:
$$ u^\varepsilon(x,t) = \int_{y \in {\mathbb R}_+} V(x,y) \, \nu^\varepsilon_{x,t}(y) \, dy. $$
Similarly, multiplying (\ref{eq:mfl_eq_eps_2_finance}) by $y$ and integrating it with respect to $y$ leads to
\begin{eqnarray}
&& \hspace{-1cm} 
\partial_t (\rho^\varepsilon \Upsilon^\varepsilon) + \nabla_x \cdot (\rho^\varepsilon {\mathcal E}^\varepsilon) = 0 , \label{eq:mw_eps}
\end{eqnarray}
with $\Upsilon^\varepsilon$ the mean wealth of $f^\varepsilon$, associated to $\nu^\varepsilon_{x,t}$ through (\ref{eq:mean_wealth}) and ${\mathcal E}^\varepsilon(x,t)$ the mean wealth velocity in configuration space, defined by:
$$ {\mathcal E}^\varepsilon(x,t) = \int_{y \in {\mathbb R}_+} y \, V(x,y) \, \nu^\varepsilon_{x,t}(y) \, dy.$$

Then, taking the limit $\varepsilon \to 0$ in (\ref{eq:mass_eps}), (\ref{eq:mw_eps}), and using that $\rho^\varepsilon \to \rho$, $\Upsilon^\varepsilon \to \Upsilon$ and $ \nu^\varepsilon_{x,t} \to M_{\Xi_{x,t}}$, we get that $u^\varepsilon(x,t) \to u (x;\Xi_{x,t})$ and ${\mathcal E}^\varepsilon(x,t) \to {\mathcal E} (x;\Xi_{x,t})$. Finally,  the limits of Eqs. (\ref{eq:mass_eps}), (\ref{eq:mw_eps}) are precisely Eqs.  (\ref{eq:mass_finance}), (\ref{eq:mw_vel_equi_finance}). 
\endproof

\medskip
Eq. (\ref{eq:mass_finance}) is a continuity equation which expresses how the number of agents in a given domain of economic neighborhood changes in time. Indeed, let $I = [a,b]$ be an interval in economic neighborhood and denote by $N_I(t)$ the number of agents in this interval at time $t$:
$$ N_I(t) = \int_{x \in I} \rho(x,t) \, dx.$$
Then, integrating (\ref{eq:mass_finance}) over $x \in I$, and using an integration by parts, we get
$$ \frac{d N_I}{dt} = - (\rho u )(b) + (\rho u )(a). $$
This equation expresses that $N_I$ varies in time due to the flux of agents leaving $I$ through point $b$ and entering $I$ through point $a$. Therefore, $\rho u $ is the flux of agents crossing an arbitrary point and $u(x;\Xi_{x,t})$  is the average velocity of the agents in economic neighborhood at this point. A similar interpretation is valid for Eq. (\ref{eq:mw_finance}), replacing agents by wealth. Indeed, the average wealth is transported in economic neighborhood with velocity ${\mathcal E}(x;\Xi_{x,t})$ as the second term of Eq. (\ref{eq:mw_finance}) expresses. 

\medskip
In the present case of a general trading interaction potential $\phi$, it is not clear if the macroscopic equations (\ref{eq:mass_finance}), (\ref{eq:mw_finance}) form a closed system of equations. For this to be true, we would need to be able to reconstruct $\Xi_{x,t}$ from the knowledge of $\rho(x,t)$ and $\Upsilon(x,t)$. This is not possible in general, unless the fixed point equation (\ref{eq:NE_fixedpoint}) has a unique solution among distributions $\nu$ of given mean wealth $\Upsilon$. Such a uniqueness statement is by no means obvious. There could exist a continuous family of solutions parametrized by a parameter belonging to a $d$-dimensional manifold. In this case, we would need $d$ additional macroscopic equations to specify how these parameters evolve in configuration space and time. The characterization of the manifold of equilibria strongly depends on the assumptions made on the trading interaction potential $\phi$ and it it not possible to make a general theory at this point. 

Here as a matter of illustration, we investigate the quadratic trading interaction case and show that the two conservation equations (\ref{eq:mass_finance}), (\ref{eq:mw_vel_equi_finance}) form a closed system of equations. This obviously follows from the fact that the twisted cost function $\Xi$ is fully determined by the knowledge of the mean wealth $\Upsilon$. By doing so, we recover the framework of \cite{During_Toscani_PhysicaA07}. However, we note that this framework is very specific and cannot be generalized simply to arbitrary trading interaction potentials.

%%%%%%%%%%%%%%%%%%%%%%%%%%%%%%%%%%%%%
%%%%%%%%%%%%%%%%%%%%%%%%%%%%%%%%%%%%%
%%%%%%%%%%%%%%%%%%%%%%%%%%%%%%%%%%%%%
%%%%%%%%%%%%%%%%%%%%%%%%%%%%%%%%%%%%%
\subsection{Quadratic trading interaction case}
\label{subsec_inhomo_quadra}

%%%%%%%%%%%%%%%%%%%%%%%%%%%%%%%%%%%%%
%%%%%%%%%%%%%%%%%%%%%%%%%%%%%%%%%%%%%
\subsubsection{LTE and conservations in the quadratic case}
\label{subsubsec_inhomo_LTE_quadra}

Here, we return to the quadratic trading interaction framework as developed in the homogeneous configuration case in section \ref{subsec:homo_quadra}. We first specify the results of section \ref{subsec_LTE} to the present case. 

Again, we make Assumption \ref{ass:alpharho=constant} for the sake of simplicity. We consider the spatially inhomogeneous kinetic equation (\ref{eq:mfl_eq_eps_2_finance}) where now, the collision operator $Q$ is given by 
\begin{eqnarray}
&&\hspace{-1cm}
Q(f) = {\mathcal Q}_{\Upsilon_{\nu_{x,t}}} (\nu_{x,\,t}) ,
\label{eq:Q_def_finance_1_BM}
\end{eqnarray}
where, for $\Upsilon \in {\mathbb R}_+$ and $\nu \in {\mathcal P}_{\mbox{\scriptsize ac}}({\mathbb R}_+) \cap C^2({\mathbb R}_+)$, ${\mathcal Q}_\Upsilon (\nu)$ is given by (\ref{eq:def_mathcalQ_BM_2}), and where $\Upsilon_\nu$ is the mean wealth of $\nu$ given by (\ref{eq:mean_wealth}). We recall that $\nu_{x,\,t}$ and $\rho(x,t)$ are given by~(\ref{eq:nu_def}),~(\ref{eq:mfl_mom_eps_finance_2}). 

The equilibria are given by the following corollary of Prop. \ref{prop:equilibria_2_BM}:

\begin{corollary}
Let $f:$ $(x,y,t) \in {\mathbb R} \times {\mathbb R}_+ \times {\mathbb R}_+ \mapsto f(x,y,t) \in {\mathbb R}_+$ such that $\forall (x,t) \in {\mathbb R} \times {\mathbb R}_+$, the function $y \mapsto f(x,y,t)$ belongs to $L^1({\mathbb R}_+) \cap C^2({\mathbb R}_+)$. Then, the following two statements are equivalent: \\
(i) $Q(f) = 0$ in the weak sense, where $Q$ is the collision operator (\ref{eq:Q_def_finance_1_BM}), \\
(ii) there exists $\rho(x,t) >0$ and $\Upsilon(x,t) >0$ such that $f$ is given by
\begin{eqnarray}
&&\hspace{-1cm}
f(x,y,t) = F_{\mbox{\scriptsize{eq}}, \, \rho(x,t), \Upsilon(x,t)} (y) :=\rho(x,t) \, M_{\Upsilon(x,t)}(y), \quad \forall (x,y,t) \in {\mathbb R} \times {\mathbb R}_+ \times {\mathbb R}_+,
\label{eq:equi_f_BM}
\end{eqnarray}
where, for any $\Upsilon \in {\mathbb R}_+$, $M_\Upsilon$ denotes the inverse gamma distribution (\ref{eq:NE_finance}). 
\label{cor:equi_BM}
\end{corollary}

Now, in the coarse-graining limit $\varepsilon \to 0$ of (\ref{eq:mfl_eq_eps_2_finance}) we get the agent and wealth density conservation equations, stated in the following theorem: 

\begin{theorem}
We consider the solution $f^\varepsilon$ to (\ref{eq:mfl_eq_eps_2_finance}) supplemented with the collision operator (\ref{eq:Q_def_finance_1_BM}). We suppose that $f^\varepsilon$ converges to a function $f$ when $\varepsilon \to 0$ smoothly, which means in particular that all derivatives of $f^\varepsilon$ converge to the corresponding derivative of~$f$. Then, formally $f$ is given by an LTE (\ref{eq:equi_f_BM}). The agent density $\rho(x,t)$ and wealth density $(\rho \Upsilon)(x,t)$ satisfy the following conservation law:
\begin{eqnarray}
&& \hspace{-1cm} 
\partial_t \rho + \nabla_x \cdot \big(\rho \, u(x;\Upsilon(x,t)) \big) = 0 , 
\label{eq:mass_finance_BM} \\
&& \hspace{-1cm} 
\partial_t (\rho \Upsilon)  + \nabla_x \cdot \big(\rho \, {\mathcal E} (x;\Upsilon(x,t)) \big) = 0 , 
\label{eq:mw_finance_BM} 
\end{eqnarray}
where, for any  $\Upsilon \in {\mathbb R}_+$, we denote by 
\begin{eqnarray}
&& \hspace{-1cm} 
u (x;\Upsilon) = \int_{y \in {\mathbb R}_+} V(x,y) \, M_\Upsilon(y) \, dy,
\label{eq:mean_vel_equi_finance_BM} \\
&& \hspace{-1cm} 
{\mathcal E} (x;\Upsilon) = \int_{y \in {\mathbb R}_+} y \, V(x,y) \, M_\Upsilon(y) \, dy.
\label{eq:mw_vel_equi_finance_BM} 
\end{eqnarray}
\label{thm:limit_epsto0_finance_BM}
\end{theorem}

We see that system (\ref{eq:mass_finance_BM}), (\ref{eq:mw_finance_BM}) forms a closed system of equations. There are two unknown scalars $\rho(x,t)$ and $\Upsilon(x,t)$ and two scalar conservation equations to determine them. This model describes the combined evolution of the density $\rho$ of agents and their average wealth $\Upsilon$ in economic neighborhood and time. It resembles a gas dynamic system. It is readily checked that the model is strictly hyperbolic (and hence, well-posed), provided the following condition is satisfied: 
$$ \big( u(\Upsilon) - \Upsilon u'(\Upsilon) + {\mathcal E}'(\Upsilon) \big)^2 > 4 u^2(\Upsilon) \big(  \frac{u}{\mathcal E} \big)'(\Upsilon), \quad \forall \Upsilon \in {\mathbb R}_+, $$
where primes denote derivatives with respect to $\Upsilon$. Specific expressions of $u(\Upsilon)$ and ${\mathcal E}(\Upsilon)$ will depend on the context. 

This type of model has previously been derived in \cite{During_Toscani_PhysicaA07}. The derivation of \cite{During_Toscani_PhysicaA07} relies on a moment method, where the zero-th and first order moments of the distribution function with respect to the wealth variable are taken, and closure using the inverse gamma distribution. Here, we wish to show a bit more, namely that there are no other independent conservation relation involved in the macroscopic system. If it were so, the limit system would be determined by more equations than unknown functions parametrizing the equilibrium distribution. This would imply that the limit problem is ill-posed and would indicate that the formal $\varepsilon \to 0$ limit cannot lead to a rigorous result. Here, we show that the zero-th and first order moments of the distribution functions are the only conserved quantities, indicating the consistency of the limit model and the possibility of converting the present formal result into a rigorous one. Proving this amounts to showing that the space of collision invariants is spanned by the functions $1$ and $y$. This is performed in the next section.

%%%%%%%%%%%%%%%%%%%%%%%%%%%%%%%%%%%%%
%%%%%%%%%%%%%%%%%%%%%%%%%%%%%%%%%%%%%
\subsubsection{Collision Invariants}
\label{subsubsec_GCI}

We recall the definition of a collision invariant (or 'CI').

\begin{definition} A function $\chi$: $y \in {\mathbb R}_+ \to \chi(y) \in {\mathbb R}$ in $C^2({\mathbb R}_+)$ is called a Collision Invariant (CI) if and only
\begin{eqnarray}
&&\hspace{-1cm}
\int_{y \in {\mathbb R}_+} Q (\nu) (y) \, \chi (y) \, dy = 0, \quad \forall \nu \in {\mathcal P}_{\mbox{\scriptsize ac}}({\mathbb R}_+) \cap C^2({\mathbb R}_+).
\label{eq:def_CI}
\end{eqnarray}
\label{def:CI_GCI}
\end{definition}

\medskip
We denote by ${\mathcal C}$ the set of CI. This is a vector space. As seen before, $\chi(y) = 1$ and $\chi(y) = y$ are CI and corresponds to the conservation of the number of agents and the conservation of wealth through trading interactions. We prove that, in the quadratic trading interaction case, ${\mathcal C}$ is actually the vector space spanned by these two functions, i.e. the dimension of ${\mathcal C}$ is equal to $2$. This means that there are no other conservation relations involved in the macroscopic limit than the conservations of the agent and wealth densities. 

\begin{proposition}
We now suppose the additional condition $\kappa > d$. The set ${\mathcal C}$ of CI is the two-dimensional vector space spanned by the functions $1$ and $y$. 
\label{prop:GCI_finance}
\end{proposition}

\medskip
\noindent
{\bf Proof.} We first transform (\ref{eq:def_CI}) into a variational formulation and then, apply Lemma \ref{lem:dual}. Let $\chi \in {\mathcal C}$. Eq. (\ref{eq:def_CI}) is equivalently written: 
\begin{eqnarray}
&&\hspace{-1cm}
\int_{y \in {\mathbb R}_+} {\mathcal Q}_{\Upsilon_\nu} (\nu) (y) \, \chi (y) \, dy = 0, \quad \forall \nu \in {\mathcal P}_{\mbox{\scriptsize ac}}({\mathbb R}_+) \cap C^2({\mathbb R}_+).
\label{eq:def_CI_2}
\end{eqnarray}
The difficulty is the nonlinearity induced by the fact that $\Upsilon$ appearing in ${\mathcal Q}$ is related to $\nu$ by $\Upsilon = \Upsilon_\nu$. We first need to break this nonlinearity. For this purpose, let $Y>0$ be fixed. Obviously, (\ref{eq:def_CI_2}) is equivalent to saying that for all $Y \in {\mathbb R}_+$, we have
\begin{eqnarray}
&&\hspace{-1cm}
\int_{y \in {\mathbb R}_+} {\mathcal Q}_Y (\nu) (y) \, \chi (y) \, dy = 0, \quad \forall \nu \in {\mathcal P}_{\mbox{\scriptsize ac}}({\mathbb R}_+) \cap C^2({\mathbb R}_+) \mbox{ such that } \Upsilon_\nu = Y.  
\label{eq:def_CI_3}
\end{eqnarray}
We first fix $Y \in {\mathbb R}_+$ and look for all function $\chi_Y$ such that (\ref{eq:def_CI_3}) holds. Let us denote by ${\mathcal C}_Y$ the space of such functions $\chi_Y$. It is also a vector space. Obviously from (\ref{eq:def_CI_3}), we have 
$$ {\mathcal C} = \bigcap_{Y \in {\mathbb R}_+} {\mathcal C}_Y. $$

Now, the constraint $\Upsilon_\nu = Y$ is a linear constraint on $\nu$ which can be written $$\int_{y \in {\mathbb R}_+} \nu (y) \, (y-Y) \, dy = 0.$$ 
Therefore, $\chi_Y \in {\mathcal C}_Y$ if and only if the following implication holds for all $\nu \in {\mathcal P}_{\mbox{\scriptsize ac}}({\mathbb R}_+) \cap C^2({\mathbb R}_+)$: 
\begin{eqnarray}
&&\hspace{-1cm}
\int_{y \in {\mathbb R}_+} \nu (y) \, (y-Y) \, dy = 0 \quad \Longrightarrow \quad \int_{y \in {\mathbb R}_+} {\mathcal Q}_Y (\nu) (y) \, \chi_Y (y) \, dy = 0.  
\label{eq:def_CI_4}
\end{eqnarray}
Both expressions to the left and right hand sides of the arrow in (\ref{eq:def_CI_4}) are linear forms of $\nu$. Therefore, by a classical duality argument, (\ref{eq:def_CI_4}) is equivalent to the existence of a real number $c$ such that for all $\nu \in {\mathcal P}_{\mbox{\scriptsize ac}}({\mathbb R}_+) \cap C^2({\mathbb R}_+)$, we have:
\begin{eqnarray}
&&\hspace{-1cm}
\int_{y \in {\mathbb R}_+} {\mathcal Q}_Y (\nu) (y) \, \chi_Y (y) \, dy = c \int_{y \in {\mathbb R}_+} \nu (y) \, (y-Y) \, dy.  
\label{eq:def_CI_5}
\end{eqnarray}
Now, we note that the operator ${\mathcal Q}_Y$ is linear. Its $L^2$ formal adjoint ${\mathcal Q}_Y^*$  is well-defined and (\ref{eq:def_CI_5}) can be written
\begin{eqnarray}
&&\hspace{-1cm}
\int_{y \in {\mathbb R}_+}  \nu (y) \, {\mathcal Q}_Y^*(\chi_Y) (y) \, dy = c \int_{y \in {\mathbb R}_+} \nu (y) \, (y-Y) \, dy.  
\label{eq:def_CI_6}
\end{eqnarray}
Since this equality is valid for all $\nu$, it is equivalent to saying that 
\begin{eqnarray}
&&\hspace{-1cm}
{\mathcal Q}_Y^*(\chi_Y) (y) \, = c  (Y-y) .
\label{eq:GCI_strong_finance}
\end{eqnarray}

Now, the derivation of the weak form of (\ref{eq:GCI_strong_finance}) is similar as in the proof of Lemma~\ref{lem:dual}~(i). We recall the definitions (\ref{eq:def_XU}), (\ref{eq:def_HU}) and (\ref{eq:def_HU0}) of ${\mathcal X}_Y$, ${\mathcal H}_Y$, ${\mathcal H}_{Y0}$. The weak form of~(\ref{eq:GCI_strong_finance}) consists in looking for $\chi_Y \in {\mathcal H}_Y$ such that:
\begin{eqnarray}
&&\hspace{-1.5cm}
\int_{y \in {\mathbb R}_+} \partial_y \chi_Y (y) \, \partial_y \sigma (y) \, y^2 \,  M_Y(y) \, dy =
c \int_{y \in {\mathbb R}_+} (y-Y) \, \sigma(y) \,  M_Y(y) \, dy
, \quad \forall \sigma \in {\mathcal H}_Y,
\label{eq:GCI_finance_weak_3}
\end{eqnarray}
By an argument already used in the proof of Lemma \ref{lem:dual} (ii), if $c =0$, the only solutions of problem (\ref{eq:GCI_finance_weak_3}) are the constants. Furthermore, by linearity, it is sufficient to look for the solutions associated  to $c$ equal to a given non-zero constant. Here we choose $c = \frac{\kappa}{d}$. Besides, by Cauchy-Schwarz inequality, we have
\begin{eqnarray*}
&&\hspace{-1.5cm}
\int_{y \in {\mathbb R}_+} (y-Y) \, \sigma(y) \,  M_Y(y) \, dy \\
&&\hspace{-1.5cm}
\leq \Big( \int_{y \in {\mathbb R}_+} (y-Y)^2 \, y^2 \, M_Y(y) \, dy \Big)^{1/2} \Big( \int_{y \in {\mathbb R}_+} \sigma^2(y) \,  M_Y(y) \, \frac{dy}{y^2} \Big)^{1/2}.
\end{eqnarray*}
Since $\sigma \in {\mathcal X}_Y$, the right-hand side of (\ref{eq:GCI_finance_weak_3}) defines a bounded linear form if and only if
$$ \int_{y \in {\mathbb R}_+} (y-Y)^2 \, y^2 \, M_Y(y) \, dy < \infty . $$
This is indeed verified if the condition $\kappa > d$ is satisfied. Additionally, thanks to (\ref{eq:mean_wealth_2}), the function $y-Y$ satisfies the solvability condition (\ref{eq:solv_cnd}). Therefore, applying Lemma \ref{lem:dual} (iii), Problem (\ref{eq:GCI_finance_weak_3}) with $c=\frac{\kappa}{d}$ admits a unique solution belonging to ${\mathcal H}_{Y0}$ which we denote by $\chi_Y^1$. We have just proved that ${\mathcal C}_Y = \mbox{Span}\{1,\chi_Y^1\}$. 

Now, we define 
$$ y_0 = \frac{\int_{y \in {\mathbb R}_+} M_Y(y) \, \frac{dy}{y}}{\int_{y \in {\mathbb R}_+} M_Y(y) \, \frac{dy}{y^2}}. $$
The function $y-y_0$ belongs to ${\mathcal H}_{Y0}$. Inserting $y-y_0$ for $\chi_Y$ into the left-hand side of variational formulation (\ref{eq:GCI_finance_weak_3}), we get
\begin{eqnarray}
\int_{y \in {\mathbb R}_+} \partial_y (y-y_0) \, \partial_y \sigma (y) \, y^2 \,  M_Y(y) \, dy &=& \int_{y \in {\mathbb R}_+} \partial_y \sigma (y) \, y^2 \,  M_Y(y) \, dy \nonumber \\
&=& - \int_{y \in {\mathbb R}_+} \sigma (y) \, \partial_y (y^2 \,  M_Y(y)) \, dy. 
\label{eq:def_CI_7}
\end{eqnarray}
But, by the definition of the equilibria $M_Y$, we have
$$ \kappa (y-Y) + d \partial_y (y^2 \,  M_Y(y)) =0. $$
Hence, (\ref{eq:def_CI_7}) leads to:
\begin{eqnarray*}
\int_{y \in {\mathbb R}_+} \partial_y (y-y_0) \, \partial_y \sigma (y) \, y^2 \,  M_Y(y) \, dy &=&  \frac{\kappa}{d}  \int_{y \in {\mathbb R}_+} \sigma (y) \, (y-Y) \, dy. 
%\label{eq:def_CI_7}
\end{eqnarray*}
Therefore, by uniqueness of the solution of (\ref{eq:GCI_finance_weak_3}) in ${\mathcal H}_{Y0}$, we have $(y-y_0) = \chi_Y^1(y)$. Since $y_0$ is a constant, we finally get that ${\mathcal C}_Y = \mbox{Span}\{1,y\}$. We see that ${\mathcal C}_Y$ does not depend on $Y$. Hence, 
$$ {\mathcal C} = \bigcap_{Y \in {\mathbb R}_+} {\mathcal C}_Y = \mbox{Span}\{1,y\}, $$
which is what needed to be proved. \endproof

%%%%%%%%%%%%%%%%%%%%%%%%%%%%%%%%%%%%%
%%%%%%%%%%%%%%%%%%%%%%%%%%%%%%%%%%%%%
%%%%%%%%%%%%%%%%%%%%%%%%%%%%%%%%%%%%%
%%%%%%%%%%%%%%%%%%%%%%%%%%%%%%%%%%%%%
%%%%%%%%%%%%%%%%%%%%%%%%%%%%%%%%%%%%%
\setcounter{equation}{0}
\section{Conclusion and perspectives}
\label{sec:conclu}

In this paper we have presented and analyzed a kinetic model of rational agents which interact by exchanging wealth and besides, evolve slowly in an economic configuration space as a result of the fast trading exchanges. Each agent bases its decisions on minimizing a cost functional. This results in a redistribution of wealth which drives the system towards a Nash equilibrium. We have considered general cost functions while the literature is mostly concerned with quadratic cost functions. On the large scales, this gives a hydrodynamic-like model for the agent and wealth densities, which leads to a closed system in the case of a quadratic cost function. There are several interesting questions left to be answered. The most important ones are the characterization of the number of Nash equilibria and of their stability in the case of a non-quadratic cost function, and the derivation of a closed system of hydrodynamic equations in this case. The existence of multiple equilibria are the indications of possible phase transitions. Such phase transitions could provide a paradigm for economic cycles and fast societal transitions which appear when a new technology emerges.

%%%%%%%%%%%%%%%%%%%%%%%%%%%%%%%%%%%%%
%%%%%%%%%%%%%%%%%%%%%%%%%%%%%%%%%%%%%
%%%%%%%%%%%%%%%%%%%%%%%%%%%%%%%%%%%%%
%%%%%%%%%%%%%%%%%%%%%%%%%%%%%%%%%%%%%
%%%%%%%%%%%%%%%%%%%%%%%%%%%%%%%%%%%%%

%%%%%%%%%%%%%%%%%%%%%%%%%%%%%%%%%%%%%
%%%%%%%%%%%%%%%%%%%%%%%%%%%%%%%%%%%%%
%%%%%%%%%%%%%%%%%%%%%%%%%%%%%%%%%%%%%
%%%%%%%%%%%%%%%%%%%%%%%%%%%%%%%%%%%%%
%%%%%%%%%%%%%%%%%%%%%%%%%%%%%%%%%%%%%
\begin{appendices}

\setcounter{equation}{0}
\section{Proof of Lemma \ref{lem:dual}}
\label{app:proof:dual}

(i) Introducing the change of variables (\ref{eq:chgvar}) into (\ref{eq:problem}) and using Green's formula, we find (\ref{eq:GCI_finance_weak_2}). Green's formula is applicable and the boundary terms disappear because of the assumptions of smoothness made on $f$ and $g$. 

\medskip
\noindent
(ii) We just let $\sigma = \varphi$ in (\ref{eq:GCI_finance_weak_2}). 

\medskip
\noindent
(iii) Taking $\sigma = \mbox{Constant}$ in (\ref{eq:GCI_finance_weak_2}), the left-hand side vanishes. Therefore, if (\ref{eq:solv_cnd}) is not satisfied, there cannot exist a solution. Supposing now that (\ref{eq:solv_cnd}) is satisfied, we can restrict the set of test functions $\sigma$ to ${\mathcal H}_{\Xi 0}$ in the weak formulation (\ref{eq:GCI_finance_weak_2}). Indeed, from $\sigma \in {\mathcal H}_{\Xi 0}$, we can construct an arbitrary test function in ${\mathcal H}_\Xi$ by simply adding a constant. But, because (\ref{eq:solv_cnd}) is satisfied, the weak formulation (\ref{eq:GCI_finance_weak_2}) is still true for this test function. Now, because of the assumed Poincar\'e inequality (\ref{eq:poincare}), the left-hand side of  (\ref{eq:GCI_finance_weak_2}) is a coercive bilinear form on ${\mathcal H}_{\Xi 0}$ while, because of the assumption that $\psi \in {\mathcal X}_\Xi$, the right-hand side is a continuous linear form on ${\mathcal H}_{\Xi 0}$. Therefore, Lax-Milgram's theorem applies and there exists a unique solution $\varphi \in {\mathcal H}_{\Xi 0}$ to problem (\ref{eq:GCI_finance_weak_2}). The most general solution is of the form $\varphi + \mbox{Constant}$ because of point (ii). This ends the proof. \endproof

\setcounter{equation}{0}
\section{Proof of Lemma \ref{lem_poincare}}
\label{app:proof:poincare}

Let $v$: $z \in {\mathbb R}_+ \mapsto v(z) \in {\mathbb R}$ such that
\begin{eqnarray}
&&\hspace{-1cm}
\int_0^\infty (|v(z)|^2 + |\partial_z v(z)|^2) \, \gamma_{\alpha,\beta}(z) \, dz < \infty,
\label{eq:space_rossignol}
\end{eqnarray}
where $\gamma_{\alpha,\beta}(z)$ is the gamma distribution defined at (\ref{eq:gamma_distrib}). Then, formula (10) of \cite{Benaim_Rossignol_arxiv} states that there exists a constant $C_{\alpha,\beta}>0$ such that
\begin{eqnarray}
&&\hspace{-1cm}
\int_0^\infty |v(z) - \bar v|^2 \, \gamma_{\alpha,\beta}(z) \, dz \leq C_{\alpha,\beta} \int_0^\infty |\partial_z v(z)|^2 \, \gamma_{\alpha,\beta}(z) \, dz, \label{eq:poincare_rossignol}
\end{eqnarray}
where
\begin{eqnarray}
&&\hspace{-1cm}
\bar v = \int_0^\infty v(z) \, \gamma_{\alpha,\beta}(z) \, dz.
\label{eq:mean_rossignol}
\end{eqnarray}
Then, we make the change of variables $z = 1/y$ in (\ref{eq:space_rossignol}), (\ref{eq:poincare_rossignol}), (\ref{eq:mean_rossignol}). We denote by $u(y) = v(z)$ and use (\ref{eq:link_gamma}). We remark that $\partial_z v(z) = - y^2 \partial_y u (y)$. Therefore, we have, denoting by $C_{\alpha,\beta} $ generic constants only depending only on $\alpha$ and $\beta$:
\begin{eqnarray*}
\int_0^\infty |u(y)|^2  \, g_{\alpha,\beta}(y) \, \frac{dy}{y^2} &=& \int_0^\infty |v(z)|^2 \, z^2 \, \gamma_{\alpha,\beta}(z) \, dz  \\
&=& C_{\alpha,\beta} \int_0^\infty |v(z)|^2 \, \gamma_{\alpha+2,\beta}(z) \, dz,
\end{eqnarray*}
and
\begin{eqnarray*}
\int_0^\infty |y^2 \, \partial_y u (y)|^2  \, g_{\alpha,\beta}(y) \, \frac{dy}{y^2} &=& \int_0^\infty |\partial_z v (z)|^2 \, z^2 \, \gamma_{\alpha,\beta}(z) \, dz  \\
&=& C_{\alpha,\beta} \int_0^\infty |\partial_z v (z)|^2 \, \gamma_{\alpha+2,\beta}(z) \, dz,
\end{eqnarray*}
and finally,
\begin{eqnarray*}
\bar u = \int_0^\infty u (y)  \, g_{\alpha,\beta}(y) \, \frac{dy}{y^2} &=& \int_0^\infty  v (z) \, z^2 \, \gamma_{\alpha,\beta}(z) \, dz  \\
&=& C_{\alpha,\beta} \int_0^\infty v (z) \,\gamma_{\alpha+2,\beta}(z) \, dz.
\end{eqnarray*}
Now, letting $(\alpha,\beta) = (\frac{\kappa +d}{d}, \frac{\kappa \Upsilon}{d})$, we notice that $v$ satisfies (\ref{eq:space_rossignol}) (with $\alpha$ shifted to $\alpha + 2$) if and only if $u \in {\mathcal H}_\Upsilon$. Furthermore, $\bar v = 0$ if and only if $u \in {\mathcal H}_{\Upsilon 0}$. Now, the Poincar\'e inequality (\ref{eq:poincare_rossignol}) (with $\alpha$ shifted to $\alpha + 2$) leads to (\ref{eq:poincare}).
\endproof

\end{appendices}

\end{document}